\documentclass[twocolumn,showpacs]{revtex4-1}

\usepackage{color}
\usepackage{hyperref}
\usepackage{graphicx}

\def\beq{\begin{equation}} \def\eeq{\end{equation}}
\def\beqa{\begin{eqnarray}} \def\eeqa{\end{eqnarray}}

\def\AA{{\it Astronomy \& Astrophysics} }
\def\AJ{{\it Ap. J.} }

\def\AJS{{\it Ap. J. Supp.} }

\def\CQG{{\it Class. Quantum Gravity} }

\def\GRG{{\it Gen. Relativity and Gravitation} }

\def\IJMP{{\it Int. J. Mod. Phys.} }

\def\MNRAS{{\it Mon. Not. R. Ast. Soc.} }

\def\PL{{\it Phys. Lett.} }
\def\PR{{\it Phys. Rev.} }

\def\PRTS{{\it Phys. Rep.} }

\def\RMP{{\it Rev. Mod. Phys.} }

\def\al{\alpha} \def\be{\beta} \def\ga{\gamma} 
\def\ep{\epsilon}   
   \def\ka{\kappa}

  \def\De{\Delta} 
\def\La{\Lambda}   
  \def\mn{{\mu\nu}} \def\cl{{\cal L}}

 \def\frac#1#2{{\textstyle{{#1}\over
{#2}}}} 
\def\lsim{\mathrel{\rlap{\lower4pt\hbox{\hskip1pt$\sim$}}
\raise1pt\hbox{$<$}}}
\def\gsim{\mathrel{\rlap{\lower4pt\hbox{\hskip1pt$\sim$}}
\raise1pt\hbox{$>$}}} \def\sqr#1#2{{\vcenter{\vbox{\hrule height.#2pt
\hbox{\vrule width.#2pt height#1pt \kern#1pt \vrule width.#2pt} \hrule
height.#2pt}}}}
\def\square{\mathchoice\sqr66\sqr66\sqr{2.1}3\sqr{1.5}3}
\def\eq#1{Eq. (\ref{#1})}
\def\nn{\nonumber}

\begin{document}

\title{Mimicking dark matter in galaxy clusters through a non-minimal gravitational coupling with matter}

\author{Orfeu Bertolami}
\email{orfeu.bertolami@fc.up.pt}
\homepage{http://web.ist.utl.pt/orfeu.bertolami}
\affiliation{Departamento de F\'{\i}sica e Astronomia, Faculdade de Ci\^encias, Universidade do Porto,\\Rua do Campo Alegre 687,
4169-007 Porto, Portugal}
\altaffiliation{Also at Instituto de Plasmas e Fus\~ao Nuclear, Instituto Superior T\'ecnico}

\author{Pedro Fraz\~ao}
\email{pedro.frazao@ist.utl.pt}

\author{Jorge P\'aramos}
\email{paramos@ist.edu}
\homepage{http://web.ist.utl.pt/jorge.paramos}
\affiliation{Instituto de Plasmas e Fus\~ao Nuclear, Instituto Superior T\'ecnico\\Av. Rovisco Pais 1, 1049-001 Lisboa, Portugal}

\date{\today}

\begin{abstract}
In this work, one shows that a specific non-minimal coupling between the scalar curvature and matter can mimic the dark matter component of galaxy clusters. For this purpose, one assesses the Abell cluster A586, a massive nearby relaxed cluster of galaxies in virial equilibrium, where direct mass estimates and strong-lensing determinations are possible. One then extends the dark matter mimicking to a large sample of galaxy clusters whose density profiles are obtained from the {\it Chandra} high quality data, also in virial equilibrium. The total density, which generally follows a cusped profile and reveals a very small baryonic component, can be effectively described within this framework. 
\end{abstract}


\maketitle

\section{Introduction}

In the past two decades, cosmology has acquired a new standing due to a wide range of observations of high precision data \cite{WMAP7,SDSS}. These observations seem to reveal that the two dominant components of the Universe are a non-baryonic form of matter, the so-called dark matter \cite{bertone}, and an exotic form of energy, dark energy \cite{copeland}. The former leads to structure formation while the latter is responsible for the present accelerated expansion of the Universe. The search for an explanation for the existence and properties of these dark components of the Universe has prompted a strong debate about their origin and nature.

The behaviour of the galaxy rotation curves and the dynamical mass in clusters of galaxies suggest the existence of an exotic form of matter, dark matter, at galactic and extra galactic scales, respectively, where the presence of this sort of matter is revealed through its gravitational effects. Notwithstanding, recent results demonstrated that dark matter may interact with ordinary matter, with the production of patterns of annihilations or decays of dark matter particles in the fluxes of cosmic rays \cite{cdm}. 

Evidence for the existence of dark matter lies in the velocity dispersion of clusters that, combined with its morphological features, leads one to conclude that the overall mass should be far greater than the visible mass. In fact, the dark matter component of these clusters is well described by dark matter models, where the interaction of dark matter with baryonic matter determines the density profile of galaxy clusters \cite{cdm_clusters}.

An alternative approach to the above mentioned problems is to consider modified theories of gravity. Amongst modified gravity theories, those that introduce higher order curvature terms in the Einstein--Hilbert action --- and, in particular, the $f(R)$ theories of gravity \cite{capozziello_DE1,faraoni_review} --- offer alternative explanations for these problems, along with other cosmological, astrophysical, and high-energy physics motivations \cite{capozziello_DE2}. 

Recently, it has been shown that, in the context of power-law theories, $f(R)\sim R^n$, the rotation curves can be explained as a curvature effect \cite{Capozziello_DM1,Capozziello_DM2}; a similar result  \cite{mimic} can be obtained in another extension of gravity that relies not only on a non-trivial curvature term, but also on a non-minimal coupling between matter and geometry \cite{coupling}. 
 
Clusters are the largest astronomical structures whose masses can be measured in a reliable form, since they constitute the largest configuration of objects that passed through gravitational relaxation and entered into virial equilibrium \cite{voit}. The measuring methods, X-rays and gravitational lensing for visible and dark matter respectively, reveal that the masses of clusters are approximately seven to ten times larger than the total combined mass of stars and hot gas in the intracluster medium (ICM). This missing mass is predominant in the inner regions of clusters, but it is also extended beyond the core radius, as defined by the gas density distribution. 

In the context of modified Newtonian dynamics \cite{MOND}, another popular (albeit incomplete \cite{MONDreview}) alternative to general relativity (GR), it is difficult to explain cluster observations without the assumption of ``real'' dark matter, because the mass of the dark component is otherwise significantly reduced but not completely removed \cite{sanders}.

The problem of clusters has recently been discussed in $f(R)$ theories, by considering a generalized version of the virial theorem. Previously discussed in the context of a possible unification and interaction of dark matter and dark energy \cite{A586_DE_DM}. Assuming a steady state, the virial theorem can be used to deduce the mean density of astrophysical objects, and therefore the corresponding total mass, such as galaxies and clusters, by observing the velocities of test particles moving around them.
The overall mass of a sample of clusters can also be estimated through corrections in the gravitational potential that emerge in the weak field limit
of these theories \cite{capozziello_modelling}. Along with these works, there were also conducted the first cluster abundance constraints on a modified gravity model,
 specifically the modified action $f(R)$ model, which predict deviations in the abundance of massive dark matter halos \cite{vikhlinin_constraints}.

In the context of the non-minimal coupling between the Ricci scalar curvature and matter \cite{coupling}, it has been shown in several works that both dark energy and dark matter can be effectively described within this framework, along with other cosmological and astrophysical implications: these include the perturbed hydrostatic equilibrium of the Sun \cite{sun}, the acceleration of the Universe \cite{coupling_dark_energy}, the galaxy rotation curves \cite{mimic}, the reheating scenario after inflation \cite{coupling_inflation}, the mimicking of a cosmological constant \cite{CC}, and the change in the gravitational potential \cite{Martins}.

In order to extend this non-minimally coupled model one tries to obtain the large mass difference of clusters through a dark matter mimicking, an extra matter component that emerges due to the presence of the non-minimal coupling, and a subsequent modification of gravity. One first explores here the scenario of the Abell cluster A586, which is a nearby massive strong-lensing cluster, with a high X-ray luminosity, related with the mass of the intracluster gas. Its morphology is well suited to the assumption of spherical symmetry, making it an ideal test bed for modifications of gravity at extra-galactic scales. One then extends this analysis to a larger set of galaxy clusters with the same features. 

This work is organized as follows. First, one shortly reviews the formalism leading to the total cluster mass and the field equations that result in the presence of the non-minimal coupling and stablishes the strategy to address the problem of mimicking the dark matter component of a cluster. One then studies the particular case of the Abell cluster A586, discusses the coupling function that leads to the dark matter mimicking, and extends the model to a larger sample of galaxy clusters. Finally, the numerical results are presented and discussed.

\section{Galaxy clusters}

\subsection{Hydrostatic equilibrium equation}
\label{hydro}

Galaxy clusters are usually considered as closed gravitational systems with spherical symmetry and in hydrostatic equilibrium when virial equilibrium is attained --- despite the fact that recent observations reveal that clusters have more evolved structures with strong interactions and dynamical activity, particularly in their inner regions \cite{innercluster1,innercluster2}. 

Under these assumptions, the structure equation can be derived from the collisionless Boltzmann equation for an isotropic system \cite{capozziello_modelling,capoziello_hidrostatic},

\beq
  {d \over dr}\left(\rho_{g} \hspace{0.1cm}\sigma_{r}^{2}\right)=-\rho_{g}  {d\Phi  \over dr}~~,
  \label{Boltzmann_equation}
\eeq

\noindent where $\Phi$ is the Newtonian gravitational potential of the cluster, $\sigma_{r}$ the mass-weighted velocity dispersion in the radial direction, and $\rho_g$ the gas-mass density. The pressure profile is related with these quantities through, $P =\sigma_{r}^{2}\rho_{g} $, and inserting in Eq. (\ref{Boltzmann_equation}) leads to $P^\prime =-\rho_{g} \Phi^\prime$, where the prime indicates differentiation with respect to the radial coordinate. The Newtonian gravitational potential $\Phi$ can be read from the Newtonian limit of the modified Einstein field equations (as addressed later), {\it i.e.} a suitably modified Poisson equation. However, this does not prevent us from formally writing the latter as the effect of the non-minimal coupling ascribed to the ``mimicked'' dark matter component, so that

\beq \nabla^2 \Phi = 4\pi G (\rho + \rho_{dm}) ~~. \label{Poissoneq}\eeq

\noindent This approach shall be discussed in more detail in the following section.

For a gas sphere, with temperature profile $T(r)$, the velocity dispersion becomes $\sigma_{r}^{2} \equiv |\sum_i <v^2> - v_i^2|  =kT /\mu m_{p}$, where $k$ is the Boltzmann constant, $\mu\approx 0.609$ the mean mass particle, and $m_{p}$ the proton mass. Introducing this expression into Eq. (\ref{Boltzmann_equation}) gives

\beqa
  {d \over dr}\left( {kT \over \mu m_{p}}\rho_{g} \right) & = & -\rho_{g} {d\Phi \over dr}~~,
\eeqa

\noindent or, equivalently,

\beqa
  -{d\Phi \over dr} & = & {kT  \over \mu m_{p}r}\left[ {d\ln\rho_{g} \over d\ln r}+{d\ln T \over d\ln r}\right]~~.
  \label{eq:Boltzmannpotential}
\eeqa

Therefore, from the models for the gas density and temperature profiles, the total mass enclosed in a sphere with a given radius $r$ can be obtained from \eq{Poissoneq},

\beqa
  M (r) & = & 4\pi \int (\rho_g + \rho_{dm}) r^2 dr = {r^2 \over G} {d\Phi \over dr} ~~,
  \label{dynamical_mass_possion}
\eeqa
leading to the usual expression for the mass profile of a spherical mass distribution in hydrostatic equilibrium
\beqa
  M (r) & = &  -{kT r\over G\mu m_{p}}\left({d\ln\rho_{g} \over d\ln r}+{d\ln T \over d\ln r}\right)~~.
  \label{dynamical_mass}
\eeqa

As already stated, this is the total mass that one should observe in a cluster and that, in general, agrees well with observations from gravitational lensing \cite{voit}. However, the baryonic component of the cluster mass, galaxies' mass in the visible spectrum and gas in X-ray spectrum, where the latter dominates, is much less than the total observed mass leading one to conclude the existence of a dark matter component that in the present work results from the presence of the non-minimal coupling. 

Hence, the equilibrium equation, Eq. (\ref{dynamical_mass}), which leads to the total mass within a cluster, can be used to derive the amount of dark matter present in a cluster of galaxies, and its spatial distribution, by the mass difference between this and the gas-mass estimates as provided by X-ray observations.

In the following, one tries to reproduce the total mass profiles of a sample of galaxy clusters, where the additional density component emerges not from an exotic form of ``dark'' matter but from a change in geometry due to the presence of the non-minimal coupling in the action.

\subsection{Cluster density profiles}

Clusters of galaxies present a velocity dispersion with a relative constant value along the radius, which reveals an implicit matter density profile of the form $\rho_{M}\propto r^{-2}$. The singular isothermal sphere is the simplest model whose density profile, $\rho_{M}=\sigma^2_{v}/2\pi G r^2$, leads to a constant and isotropic velocity
dispersion $\sigma_{v}$ along the clusters' profile. 

However, numerical fits to the observational data show that the density profile of clusters are flatter than the isothermal one at small radii and steeper at larger radii. A general form replicating this behaviour is given by the cusped density profile,

\beqa
\rho_{M}&=\rho_0 &\left( {r \over r_s} \right)^{-p}\left(1+{r \over r_s}\right)^{p-q}~~,
\label{cusped_profile}
\eeqa

\noindent where the parameters $p$ and $q$ determine the respective slopes of the power-law density profile in the inner and outer regions, and $r_s$ is the so-called core radius, signaling the steepening of the profile. These parameters usually present values within the ranges $1\lesssim p \lesssim 1.5$ and $2.5\lesssim q \lesssim 3$, for models such as the Navarro-Frenk-White (NFW) profile ($p=1$ and $q=3$), the Moore profile ($p=1.5$ and $q=3$), and the Rasia profile ($p=1$ and $q=2.5$). In fact, one of the important results of hierarchical CDM models is that the density profile of cold dark matter halos is described by the NFW model \cite{NFW}.
 
The more relevant baryonic component is the ICM in the form of hot gas, which emits observable X-ray radiation. In general, the observed surface-brightness profiles of clusters that reveal the underlying visible matter profile of this intracluster gas are well described by the so-called $\beta$ model:

\beq
\rho_g(r)=\rho_{g0}\left(1+{r^2 \over r^2_c}\right)^{-3\beta/2}~~,
\label{visible_profile}
\eeq

\noindent where $r_c$ is the core radius, which has a specific value for each cluster \cite{betamodel}.

\subsection{Abell Cluster A586}

In this cluster, the emission measured profile is well described by the $\beta$ model,  Eq. (\ref{visible_profile}), and, from a least-squares fit of two different sets of observational data, the obtained values for the free parameters are $\beta=0.518\pm 0.006$ and $r_c=67\pm 2 \; h_{70}^{-1}\,\rm{kpc}$ \cite{A586_data}.  The ICM particle number density profile is given directly by the analytic fit for the projected emission measure profile, Eq. (\ref{visible_profile}), and is easily converted into the gas density.  

The deep potential well of the cluster compresses the ICM gas to X-ray emitting temperatures, such that the gas temperature can be inferred from the derived spectrum. For a polytropic temperature model \cite{A586_data},

\beqa
T(r)&=&T_0 \left(1+{r^2 \over r_c^2}\right)^{-3\beta(\gamma-1)/2}~~,
\label{temperature_profile}
\eeqa

\noindent where $r_c$ and $\beta$ are parameters obtained in the fit for the visible mass density, and a least-squares fit leads to the values $T_0=8.99\pm 0.34\;\rm{keV}$,  $\gamma=1.10\pm 0.03$ \cite{A586_data}. Naturally, setting $\gamma=1$ leads to the isothermal temperature profiles.

The total mass follows from the hydrostatic equilibrium equation, Eq. (\ref{dynamical_mass}), using the previous expressions for the visible density and
temperature profiles

\beqa
  M(r)&=& M_0 \left({r \over r_c}\right)^3\left(1+{r^2 \over r^2_c }\right)^{-1-{3\over 2}(\gamma-1)\beta}~~,
\label{mass_profile}
\eeqa

\noindent with

\beq M_0 = {3 k T_0 \beta \gamma r_c \over G \mu m_p}~~. \eeq

From the relation for a spherical mass distribution, 
\beq 
{d M \over dr}=4\pi r^2 \rho(r)~~,
\eeq

\noindent one may approximate the dark matter density by the total density profile using Eq. (\ref{mass_profile}) in the last expression and solving for $\rho(r)$,

\beqa \label{dm_profile}
&& \rho_{dm}(r) \approx 
 \rho_{dm0}\left(1+\ep{r^2\over r_c^2}\right)\left(1+{r^2\over r_c^2}\right)^{-2\ep}~~.
\eeqa

\noindent where 
 
\beqa
\rho_{dm0}&=&{3 k T_0 \beta \gamma \over 4\pi G \mu m_p r_c^2} = {M_0 \over 4\pi r_c^3} \label{defdm0}~~,
\eeqa 

\noindent and one defines $\ep \equiv [1 + 3(1-\ga)]\be$, for convenience.

Finally, one introduces the virial radius $r_V$, which signals the distance where the density of the cluster approaches a multiple $\al$ of the critical density of the universe  $\rho_c=3H_0^2/8\pi G$, where $H_0$ is the Hubble expansion rate at the present. Thus, from $\rho_{dm}(r_V) = \al \rho_c$, one may estimate the virial radius from the relation Eq. (\ref{dm_profile}),

\beqa
{r_V\over r_{c}}& \approx & \left({\rho_{dm0}\over \rho_c} {\epsilon\over\alpha}\right)^{1 /(2-\epsilon +\beta)}~~,
\eeqa	

\noindent which, for $\rho_{dm}\approx 200\rho_c$, with $\alpha=200$, leads to a virial radius of $r_V \approx 1.497\;{\rm Mpc}$.

\section{Non-minimal Coupled Model}

\subsection{Field equations}

Recently, a generalization of the $f(R)$ modified theories of gravity was considered that further extends the presence of curvature invariant terms in the Einstein--Hilbert action by non-minimally coupling the scalar curvature with matter \cite{coupling}. The action is obtained by adding an additional $f_2(R)$ term, dependent on the scalar curvature $R$ and coupled with the matter Lagrangian ${\cal L}_m$, to the usual metric theories of gravity, leading to \cite{coupling}

\beq 
S = \int \left[ {1 \over 2}f_1(R) + \left[1+f_2(R)\right] \mathcal{L}_m \right] \sqrt{-g} d^4 x~~, 
\label{model}
\eeq

\noindent where $f_i(R)$ (with $i=1,2$) are arbitrary functions of the scalar curvature and $g$ is the metric determinant. Setting $f_2(R)=0$ one obtains the usual $f(R)$ theories of gravity, and GR is recovered if one takes the linear function $f_1(R)=2\kappa R$,  where $\kappa=c^4/16\pi G$.

Variation with respect to the metric, $g_{\mu\nu}$, leads to the field equations,

\beqa
  &&\left(F_{1}+2  F_{2}\mathcal{L}_{m}\right)R_{\mu\nu}-{1 \over 2}f_{1}g_{\mu\nu}=\nn\\
  &&\quad=\left(\square_{\mu\nu}-g_{\mu\nu}\square\right)\left(F_{1}+2  F_{2}\mathcal{L}_{m}\right)+\left(1+f_{2}\right)T_{\mu\nu}~~,
  \label{field_equations}
\eeqa
where $T_{\mu\nu}$ is the energy-momentum tensor. The respective trace equation is given by

\beqa
  &&\left(F_{1}+2  F_{2}\mathcal{L}_{m}\right)R-2f_1=\nn\\
  && \quad=-3\square\left(F_{1}+2  F_{2}\mathcal{L}_{m}\right)+\left(1+f_{2}\right)T~~.
  \label{field_equations_trace}
\eeqa

\subsection{Non-conservation of the energy-momentum tensor}

\label{subsectionnon}

From the covariant derivative of the field equations, Eqs. (\ref{field_equations}), along with the Bianchi identities, $\nabla_\mu G^\mn = 0$, one encounters the covariant non-conservation of the energy-momentum tensor \cite{coupling},

\beqa
\nabla_\mu T^\mn &=& \left( g^\mn \mathcal{L}_m - T^\mn \right) \nabla_\mu \log \left(1+f_2 \right) ~~,
\label{non-cons} 
\eeqa

\noindent which can be regarded as an energy-momentum exchange between matter and geometry. The usual covariant conservation for the energy-momentum tensor is naturally obtained when the non-minimal coupling vanishes, $f_2(R)=0$.

This non-conservation implies that a massive particle in the absence of forces will not describe a geodesic curve. Strong variations on the extra term in the last equation can imply a violation of the equivalence principle, leading to a possible way of testing and setting bounds on the coupling functions, $f_2(R)$. Contrary to the usual Jordan--Brans--Dicke theories, it is impossible to perform a conformal transformation to the Einstein frame such that the coupling disappears and the covariant conservation of the energy-momentum tensor is recovered for all matter forms \cite{Sotiriou1}.

Indeed (following Ref. \cite{Sotiriou1}), using the conformal transformation

\beq g_\mn \rightarrow \tilde{g}_\mn = f_2 g_\mn ~~, \eeq

\noindent one obtains the covariant conservation law

\beq \tilde{\nabla}_\mu\tilde{T}^\mn = 0 ~~, \eeq

\noindent only if $\tilde{T}^\mn = f_2^{-2} T^\mn$ and $ 2\cl = T $.
From the energy-momentum tensor of a perfect fluid, $T_\mn = (\rho + p)u_\mu u_\nu + p g_\mn$, where $\rho$ is the energy density, $p$ the pressure and $U_\mu$ the four-velocity (with $U_\mu U^\mu = -1$), the respective trace is  $T = 3p -\rho$, such that one may write,

\beqa
2\cl = 3p-\rho \quad\rightarrow \quad p=(2\cl+\rho)/3~~.
\eeqa   

However, one obtains different equations of state depending on the chosen Lagrangian density $\cl$. For instance, if $\cl = -\rho$ \cite{fluid}, one has the state equation  $p = -\rho/3$, a perfect fluid with negative pressure, even though not a cosmological constant for which $p_\La = -\rho_\La$; for $\cl = p$, then $p = \rho$, which is the equation of state for ultra-stiff matter. Therefore, this non-conservation is, in fact, a fundamental property of the model, Eq. (\ref{model}), meaning that even under a suitable conformal transformation the energy--momentum tensor is not conserved for all different types of matter. 

Assuming that matter is described by a perfect fluid, the non-conservation, Eq. (\ref{non-cons}), results in an extra force of the form

\beqa 
f^{\mu}&=&{1 \over \rho +p} \Bigg[\left({\cal L}_m+p\right)\nabla_\nu \log \left(1+f_2\right)+\nabla_\nu p \Bigg] h^\mn,
\label{extra-force}
\eeqa

\noindent which has units of an acceleration, and where the projection operator is given by $h_\mn = g_{\mn}-U_{\mu}U_{\nu}$, 
such that the extra force is orthogonal to the four-velocity, $h_{\mn}U^{\mu }=0$.

\subsection{Mimicking the dark matter component}
\label{mimicking1}

In order to test the effect of the non-minimal coupling, one focuses on possible deviations from GR due to the 
presence of the coupling by considering the linear function $f_{1}(R)=2\kappa R$; following Ref. \cite{mimic}, one adopts a power-law non-minimal coupling 

\beq f_{2}\left(R\right)=\left({ R \over R_{n}} \right)^{n} ~~, \label{powerlawcoupling} \eeq

\noindent where $R_n$ is a characteristic curvature.

Considering a pressureless perfect fluid, $T_{\mu\nu}=\rho U_{\mu}U_{\nu}$,
and the respective trace equation, $T=-\rho$, along with the Lagrangian density, ${\cal L}_{m}=-\rho$ (a choice discussed in depth in Ref. \cite{fluid}), the field Eqs. (\ref{field_equations}) become

\beqa
   &&\left[1-{n \over \kappa}\left({R \over R_{n}}\right)^{n} {\rho \over R}\right]R_{\mu\nu}- {1 \over 2}Rg_{\mu\nu}=\nn\\
   &&\quad= {n \over \kappa}\left(g_{\mu\nu}\square-\square_{\mu\nu}\right)\left[\left({R \over R_{n}}\right)^{n} {\rho \over R}\right]\nn\\
   &&\qquad + {1 \over 2\kappa}\left[1+\left({R \over R_{n}}\right)^{n}\right]\rho U_{\mu}U_{\nu}~~,
  \label{first_equation}
\eeqa

\noindent and the respective trace, Eq. (\ref{field_equations_trace}), reads

\beq
  R={1 \over 2\kappa}\left[1+(1-2n)\left({R \over R_{n}}\right)^{n}\right]\rho-{3n \over \kappa}\square\left[\left( {R \over R_{n}}\right)^{n} {\rho \over R}\right]~~.
\label{second_equation}
\eeq

If one considers a strong coupling $(R/R_n)^n \gg 1$, the above has an implicit solution given by the vanishing of the derivative term in the {\it r.h.s.},

\beq
  {R \over R_{n}} = {\rho_{dm} \over \rho_n} \approx \left[(1-2n) {\rho \over \rho_n}\right]^{1/(1-n)}~~,
  \label{implicit}
\eeq

\noindent where one introduces the characteristic density, $\rho_n\equiv 2\kappa R_{n}$. One dubs this a ``static'' solution, not because it is time-independent (which it is, if one assumes a time--independent visible matter density $\rho(r)$), but in opposition to a more evolved, ``dynamical'' solution that would arise if this second term did not vanish.

Indeed, it was shown that (see Ref. \cite{mimic})  the most general solution of the differential equation corresponds to a dominance of the gradient terms on the {\it r.h.s.} of this equation: however, it was found that this ``dynamical'' solution exhibits very small oscillations around the ``static'' one (with a very short-wavelength providing the dominance of the gradient term). Thus, the scaling law $\rho_{dm(n)} \sim \rho^{1/(1-n)}$ remains valid even if the gradient term dominates the dynamics.

Furthermore, it was shown that, although the above analytical treatment remains valid, this implies that the non-minimal coupling is actually perturbative, {\it i.e.} $f_2 = (R/R_n)^n \ll 1$. This will be used in the following section to avoid the destabilizing effect of the extra--force \eq{extra-force} on closed orbits.

As already seen in a previous work on the mimicking of the dark matter component in galaxies \cite{mimic}, considering the tensor character of the field Eqs. (\ref{field_equations}) leads to the following relation between the energy density and pressure of the mimicked dark matter component for a power-law coupling function \cite{mimic}:

\beqa
  \label{dm_density}
  \rho_{dm} & = & {1-n\over 1-4n}2\kappa R~~,\\
  p_{dm} & = & {n \over 1-4n}2\kappa R~~,
\eeqa

\noindent which, combined with Eq. (\ref{implicit}), leads to a relation between the visible and dark matter density profiles, 

\beq
{\rho_{dm} \over \rho_n}={1-n \over 1-4n}\left[(1-2n) {\rho \over \rho_n}\right]^{1/(1-n)}~~.
\label{dm_visible2}
\eeq

\noindent The above clearly shows that one may obtain a ``dark'' component from the visible matter density profile, with a change of slope given by the scaling exponent $1/(1-n)$, {\it i.e.}, $\rho_{dm}\sim \rho^{1/(1-n)}$. Conversely, if $\rho_{dm}\sim r^{-m^\prime}$ and $\rho\sim r^{-m}$, one has

\beq
{1\over 1-n}={m^\prime \over m}\quad \rightarrow \quad n=1-{m\over m^\prime}~~,
\label{n-estimative}
\eeq

\noindent which leads to an estimate for the exponent $n$ of the power-law coupling function, Eq. (\ref{powerlawcoupling}) based on the knowledge of the visible and dark matter density profiles. 

For convenience, Eq. (\ref{second_equation}) can be rewritten in a dimensionless form, by defining the rescaled quantities \cite{mimic},

\beqa
y & \equiv & {r_n \over r} ~~, \label{var1}\\
\theta & \equiv & \left({\rho \over \rho_n}\right)^{1/(1-n)}~~,\label{var2}\\
\varrho & \equiv & {2\kappa R \over \rho_n\theta}={1 \over \theta} {R \over R_{n}}~~,\label{var3}
\eeqa

\noindent where one defines the length scale, $r_{n}=1/\sqrt{R_{n}}$. The resulting dimensionless form of the trace Eq. (\ref{second_equation}) is

\beqa
  \varrho&=&\theta^{-n}+(1-2n)\varrho^{n}
- {6n \over \theta}\bar{\square}\left(\varrho^{n-1}\right)~~,
\label{rhovar}
\eeqa

\noindent where $\bar{\square}=y^{4}d^{2}/d^{2}y$. The ``static'' equivalent solution, \eq{implicit}, is thus, $\varrho = (1-2n)^{1 /( 1-n )}$, as may easily be checked. 

Hence, by considering a known visible matter density profile $\rho$ (which in a cluster is approximately equal to the ICM gas density, $\rho\approx \rho_g$), one may solve the differential equation Eq. (\ref{rhovar}) and thus read the resulting mimicked dark matter profile through Eqs. (\ref{var3}) and (\ref{dm_density}). This component can account for the large mass difference in the cluster total density profile, as will be discussed in Sec. \ref{results} --- while the perturbative nature of the non-minimal coupling ensures that no deviation from the geodesic motion occurs.

\subsection{Modified Poisson equation}

Before further exploring the mimicking mechanism offered by the scaling law $\rho_{dm(n)} \sim \rho^{1/(1-n)}$, one addresses an issue mentioned in Sec. \ref{hydro}, namely, that the ensued modification of the Poisson equation due to the presence of a non-minimal coupling would imply that the derivation of \eq{dynamical_mass} is flawed.

One begins by recalling that the Poisson equation is obtained from the Newtonian limit of the metric,

\beq ds^2 = -[1 + 2\Phi(r)]dt^2 + \delta_{ij}dx^idx^j ~~.\label{metric}\eeq

\noindent Inserting the trace of the Einstein field equation (for non-relativistic particles with $p=0$) into the $00$ component of the latter, yields the well-known result

\beq R = -{T \over 2\ka}= {\rho \over 2\ka} \rightarrow R_{00} \approx \nabla^2 \Phi = {1 \over 4\ka} \rho~~.\eeq

\noindent Since the above metric yields $G_{00}=0$, one may instead resort directly to the scalar curvature, to obtain the same result:

\beq R = 2R_{00} \approx 2 \nabla^2 \Phi = {1 \over 2\ka}\rho~~. \eeq

\noindent Since this study focuses particularly on the trace \eq{second_equation}, the latter form is more suitable --- but both approaches are completely equivalent, as they depend only on the assumed metric \eq{metric}, not the underlying gravity model \eq{model}.

As before, \eq{second_equation} may be written as $R = (\rho+\rho_{dm})/2\ka$, with 

\beq \rho_{dm} = (1-2n)\left({R \over R_{n}}\right)^{n}\rho-6n \square\left[\left( {R \over R_{n}}\right)^{n} {\rho \over R}\right]~~. \eeq

\noindent Thus, the above shows that one trivially obtains the same form for the Poisson equation, \eq{Poissoneq}.

One can argue that this rewriting obfuscates the effect of the non-minimal coupling, as its {\it r.h.s.} also involves the scalar curvature $R \approx 2 \nabla^2\Phi$. This, however, misses the main point of this study: one is not aiming at fully determining the Newtonian gravitational potential $\Phi$ and the interconnected equilibrium density profile $\rho$ --- something that would require the derivation of a modified hydrostatic equilibrium equation, as followed in Ref. \cite{sun}.
Instead, the focus of this study (cf. the previous work addressing galactic dark matter \cite{mimic}) is to establish a relation between the observed gas density profile and what is usually interpreted as dark matter, not to explore why the former adopts a particular configuration: the mimicking mechanism proposed here acts as a ``translator'' between visible and dark matter.

\subsection{Non-geodesic motion}
\label{nongeodesicalmotion}

Before dwelling into the analytical and numerical details of the scenario under scrutiny, one must first consider a possible issue: the presence of an extra-force, as given by Eq. (\ref{extra-force}). Considering a pressureless perfect fluid along with the Lagrangian, ${\cal L}_m=-\rho$, the same equation reduces to

\beqa
f^\mu&=&-\nabla_\nu\log \left(1+f_2\right) h^{\mn} ~~.
\eeqa
For spherical symmetry, the radial component is 

\beqa
f^r&=&-\partial_r\log\left(  1+f_2\right) = - {F_2(R)\over 1+f_2(R)} {R^\prime}~~,
\eeqa

\noindent where the prime denotes a derivative with respect to the radial coordinate. 
Considering a power-law coupling of the form of \eq{powerlawcoupling}, one has

\beqa
f^r &=& -n {\left(R/R_n\right)^{n}\over 1+\left(R/R_n\right)^n} {R^\prime\over R}~~.
\label{f_r}
\eeqa

\noindent If the curvature $R$ decreases along the radius, a negative exponent $n$ will produce an inward binding force, and a positive one an outward force, which could break the stability of an orbit. 

As already seen in a previous work \cite{mimic}, different solutions for the differential equation, Eq. (\ref{rhovar}), lead to different behaviours of this extra-force, depending on the relevance of the coupling function. If it dominates, $\left(R/R_n\right)^n \gg 1$, then the force becomes

\beq f_r\approx -n{R^\prime \over R} ~~. \eeq

\noindent On the contrary, a perturbative regime leads to a suppression of the expression above by a factor $(R/R_n)^n \ll 1 $, 

\beq f_r\approx -n\left({R \over R_n}\right)^n{R^\prime \over R} ~~.\eeq

The ``static'' solution given by \eq{implicit} is derived from the dominant condition $f_2(R) \gg 1$; taking only the outer slope behaviour of cusped density profiles for the dark matter component, $\rho_{dm} \propto r^{-m}$, and a positive exponent $n$ (as will be used in the following section), one concludes that the extra force has an outward direction,

\beq f_r\approx -n{R^\prime \over R} \sim -n{\rho'_{dm} \over \rho_{dm}} = {nm \over r}~~. \eeq

For $n$ and $m$ of order unity, one obtains a force much greater than the Newtonian counterpart, $f_N = GM(r) / r^2$, thus destabilizing the orbital motion of test particles. In the inner region, $r \ll r_c$, one has an approximately constant dark matter density, $\rho_{dm} \approx \rho_{dm0}$, yielding a negligible extra force.

However, as discussed in the previous section, one finds that the ``dynamical'' solution to \eq{first_equation} must be considered, as the gradient term on the {\it r.h.s.} of the latter dominates: this solution is essentially proportional to \eq{implicit} (disregarding small, short-wavelength oscillations). Thus, for a fixed $R_n$, it is then clear that $f_2(R) $ increases for lower values of a positive exponent $n$; thus, one expects that there is a lower bound on $n$, such that the condition $f^r < f_N$ fails due to the insufficient suppression of the extra force $f^r$.

In order to assess the value of this lower bound, one first uses Eqs. (\ref{mass_profile}) and (\ref{dm_profile}) to write the following approximations, valid in the large radius region $r \gg r_c$:

\beqa \label{radial_scaling_0}
    M(r) \sim M_0 \left({r \over r_c}\right)^{1+\ep - \be}  ~~ , ~~~~ \rho_{dm}\sim\rho_{dm0}\ep \left({r \over r_c}\right)^{-2+\ep -\be} ~~.
\eeqa

\noindent Using the above, one may write

\beqa f_r &\approx& -n\left({R \over R_n}\right)^n{R^\prime \over R} \sim -n\left({\rho_{dm} \over \rho_n}\right)^n{\rho'_{dm} \over \rho_{dm}} \sim \\  && (2-\ep+\be){n\over r_c} \left({\rho_{dm0}  \over \rho_n} \ep \right)^n \left({r \over r_c}\right)^{(-2+\ep-\be)n-1}~~, \nonumber \eeqa

\noindent so that, writing $\rho_{dm0} = 2\ka/r_0^2$, the perturbative condition $ f^r \ll f_N = GM(r)/r^2$ reads

\beq \sqrt{2n(2-\ep+\be)} { r_0 \over r_c} \left(\sqrt{\ep}{r_n \over r_0} \right)^{n} \left({r \over r_c}\right)^{(\ep-\be){n-1\over 2}-n}  \ll 1~~.   \eeq

One now assumes that $-2n+(\ep-\be)(n-1) > 0$ (as is the case with the Abell cluster A586, discussed in the following section). Thus, the relevance of the extra force decreases with the distance from the cluster's center, and one may saturate the {\it l.h.s.} of the above equation by setting $r \sim r_c$, yielding

\beq b(r_n,n) \equiv \sqrt{2n(2-\ep+\be)} \left(\sqrt{\ep}{r_n \over r_0} \right)^n {r_0 \over r_c} \ll  1 ~~. \label{lower_bound}  \eeq

Finally, notice that, for a positive exponent $n$, the perturbative condition $f_2 \ll 1$ translates into

\beq \left({R \over R_n}\right)^n \ll 1 \rightarrow \rho_{dm0} \ll 2\ka R_n \rightarrow r_0 \gg r_n~~.  \eeq

\noindent Thus, a lower exponent $n$ will increase the {\it l.h.s.}, as expected. In the following section, this lower bound for $n$ will be obtained for the Abell cluster A586.

\section{The case of the Abell A586 cluster}

In this section, one tries to describe the mass difference of the Abell cluster A586. This particular cluster is a nearby massive strong-lensing cluster and has a high X-ray luminosity, directly proportional to the mass of the intracluster X-ray emitting gas. These features imply a stringent bound for the two measuring methods for its density profile: strong lensing and hydrostatic equilibrium estimatives \cite{A586_data}. The assumption of spherical symmetry is an accurate hypothesis for its morphology \cite{oguri}. The Abell cluster A586 is also a relaxed cluster, as pointed out in Ref. \cite{A586_data}.

\subsection{Density Profiles}

Recalling Eqs. (\ref{visible_profile}) and (\ref{radial_scaling_0}), one writes the approximations valid in the large radius regime $r \gg r_c$,

\beq \rho_g\sim \left({r\over r_c}\right)^{-3\beta} \quad , \quad \rho_{dm}\sim\left({r\over r_c}\right)^{-2-3(\gamma-1)\beta} ~~. \label{radial_scaling} \eeq

\noindent Therefore, from the scaling given in Eq. (\ref{dm_visible2}), one obtains the relation $\rho_{dm}\sim\rho^{1/1-n}$, allowing for an estimate, through Eq. (\ref{n-estimative}), for the exponent $n$:

\beq
\label{expected} n={2+3(\gamma-2)\beta\over2+3(\gamma-1)\beta}\approx 0.279~~,
\eeq

\noindent using $\beta=0.518$ and $\gamma=1.10$ \cite{A586_data}. Thus, for the isothermal case ($\gamma=1$) 

\beqa
n=1-{3\beta\over2}(2-\gamma)\approx 0.223~~.
\eeqa
 
\noindent Hence, the results arising from the non-minimal coupling are somewhat similar whether one considers an isothermal ($\gamma = 1$) or a polytropic temperature profile.

\subsection{Numerical results}

\label{results}

In order to mimic the dark matter component of this cluster, one numerically solves the differential Eq. (\ref{rhovar}), for an assumed gas density profile, as given by Eq. (\ref{visible_profile}), and varies the parameters $R_n$ and $n$ in order to obtain a best fit between the derived mimicked dark matter density $\rho_{total} - \rho_{gas}$ and the observed curve. The considered range for the numerical integration lies between $r_i=10r_{s}$, the Scharwzschild radius of the cluster, and a sufficiently large radius, larger than the observed size of the cluster.

As a measure of the quality one tries to minimize the quantity $\sigma_{\log}^2$, given by

\beqa
\sigma_{\log}^2=\sum_{i=1}^N {\left(\log\bar{\rho}_i-\log\rho_i\right)^2 \over \left(\log\rho_i\right)^2}~~,
\label{sigma2}
\eeqa

\noindent where the sum is performed over a large number of equal spaced radii $r_i$, the density $\rho_i$ corresponds to the dark matter profile obtained from the hydrostatic equilibrium estimatives for the radius $r_i$, and $\bar{\rho}_i$ is the value for the obtained mimicked profile in the same radius. One also presents the deviation of the final total mass of the dark matter component from the quantity 

\beq
{\Delta M\over M}={|M_{mim}-M|\over M} ~~,
\label{mass_deviation}
\eeq

\noindent where $M=M_{dm}$ is the amount of dark matter. 

\begin{figure}[bp]
  \centering
  \includegraphics[width=\columnwidth]{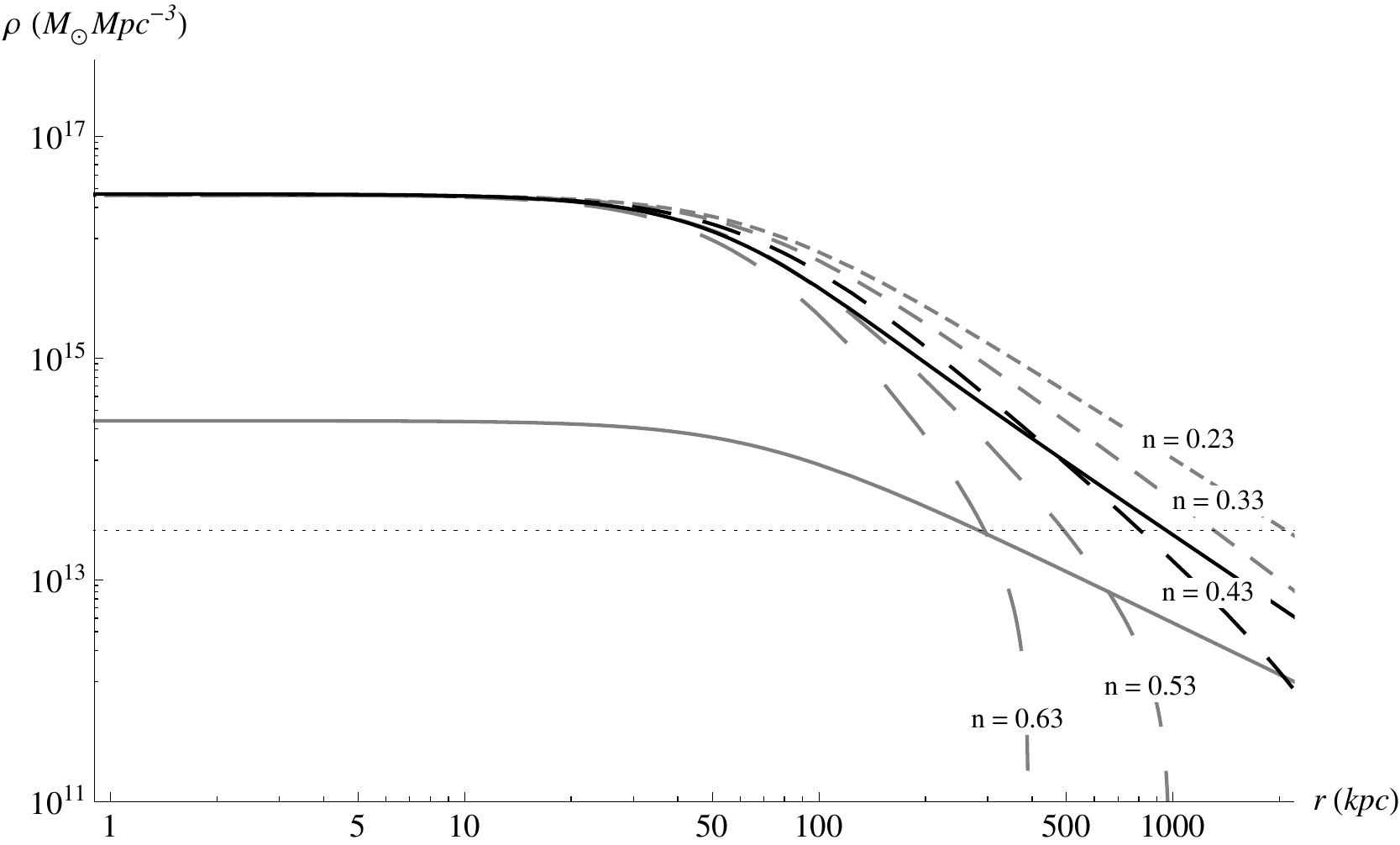}
  \caption{Mass-density profiles for the visible matter component (gray), the dark matter density profile (dark), and the mimicked density profiles (dashed) for different values of $n$, where the dashing length increases with $n$. The best fit is obtained for $n=0.43$ (dark dashed). The horizontal dotted line corresponds to $\rho=200 \rho_c$.}
  \label{densities}
\end{figure}

\begin{figure}[bp]
  \centering
  \includegraphics[width=\columnwidth]{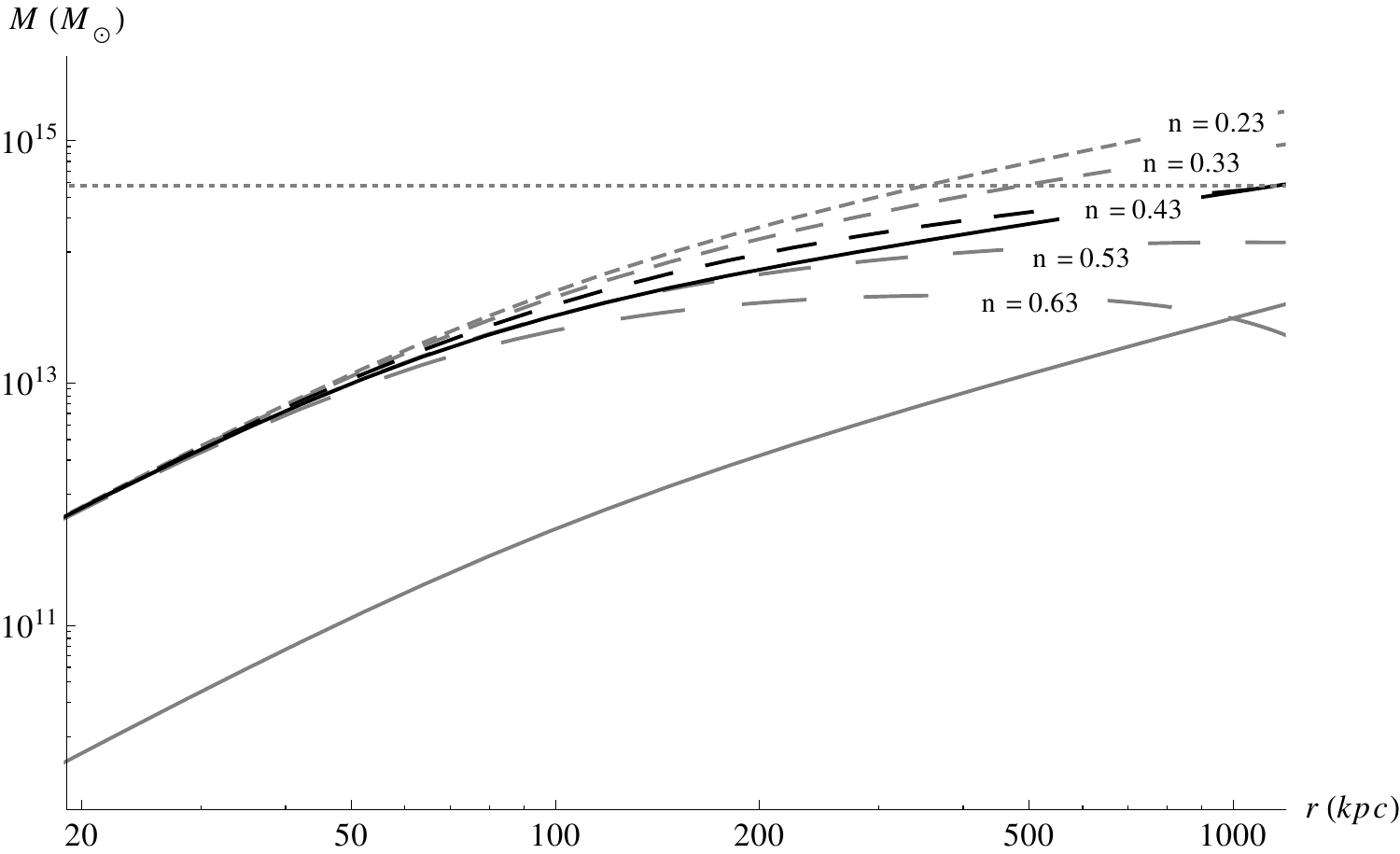}
  \caption{Mass radial profile, resulting from the integration of the mass--density profile. The dotted line is for the dynamical mass of $M=4.25\times 10^{14}\;M_{\odot}$ (polytropic case) \cite{A586_data}.}
  \label{masses}
\end{figure}

The results of the numerical integration are shown in Fig. (\ref{densities}) and the resulting integrated mass radial profile, $M(r)=\int_0^r\rho(r^\prime)dr^\prime$, in Fig. (\ref{masses}), for the values of $n=0.23$, $0.33$, $0.43$, $0.53$, and $0.63$, with a  best fit attained for $n=0.43$; $r_n = 10^{-2}~{\rm pc} $ is kept constant, as its value does not change the mimicked dark matter density profile (as discussed below). 
The best fit is given by the minimization of the relative difference between the obtained mimicked total mass and the expected one from the hydrostatic estimative, Eqs. (\ref{dynamical_mass}) and (\ref{mass_profile}), with values shown in Table (\ref{mass_difference}). The relative mass difference is evaluated at the virial radius of the Abell cluster A586,  $r_V=1.165\;{\rm Mpc}$ \cite{A586_data}.
 
As one can see, these results show that, in the outer regions of the cluster, the density decreases with the value of $n$ along with a decrease in the total mass for the cluster. The best fit $n=0.43$ is the one that yields a mimicked profile, which minimizes the mass difference to only $6.94\%$ at $r_V=1.165\;\
{\rm Mpc}$, the virial radius for the density profile derived from Eqs. (\ref{visible_profile}) and (\ref{mass_profile}). In fact, in Fig. (\ref{masses}), one can see that the values in the range, $n<0.43$, lead to a dynamical mass larger than the upper limit set by the model, $M=4.25\times 10^{14}M_\odot$.

Notice that this best fit $n = 0.43$ deviates from the expected value $n = 0.279$, computed in Eq. (\ref{expected}). As the curves for $n=0.33$ and $n= 0.23$ show, this stems from the fact that the mimicked dark matter density arising from the expected value is unable to follow the observed profile. Indeed, while $n = 0.279$ yields the same slope as the observed total mass density, it leads to a right-shifted curve, giving rise to a much greater total mass for the cluster.

\begin{table} \vspace{1.4cm}
 \centering
\begin{tabular}{c|cc|cc} 
  n & $r_V ({\rm Mpc})$& $M_V({\rm M_\odot})$ & $\Delta M/M$ & $\sigma_{\log}^2$\\\hline
  0.23 & 2.089 & $3.08\times10^{15}$  & 3.088 & 0.304\\ 
  0.33 & 1.309 & $9.99\times10^{14}$  & 1.240 & 0.108\\
  0.43 & 0.810 & $3.42\times10^{14}$  & 0.069 & 0.009\\
  0.53 & 0.493 & $1.28\times10^{14}$  & 0.561 & 0.694\\
  0.63 & 0.294 & $5.13\times10^{13}$  & 0.838 & 1.664
\end{tabular}
\caption{Virial radius and respective mass for $\rho_{dm}\approx 200\rho_c$. Mass differences for different values of the $n$ evaluated at $r=1.165\;{\rm Mpc}$.}
\label{mass_difference}
\end{table}

\begin{figure}[tbp]
  \vspace{0.4 cm}
  \centering
  \includegraphics[width=\columnwidth]{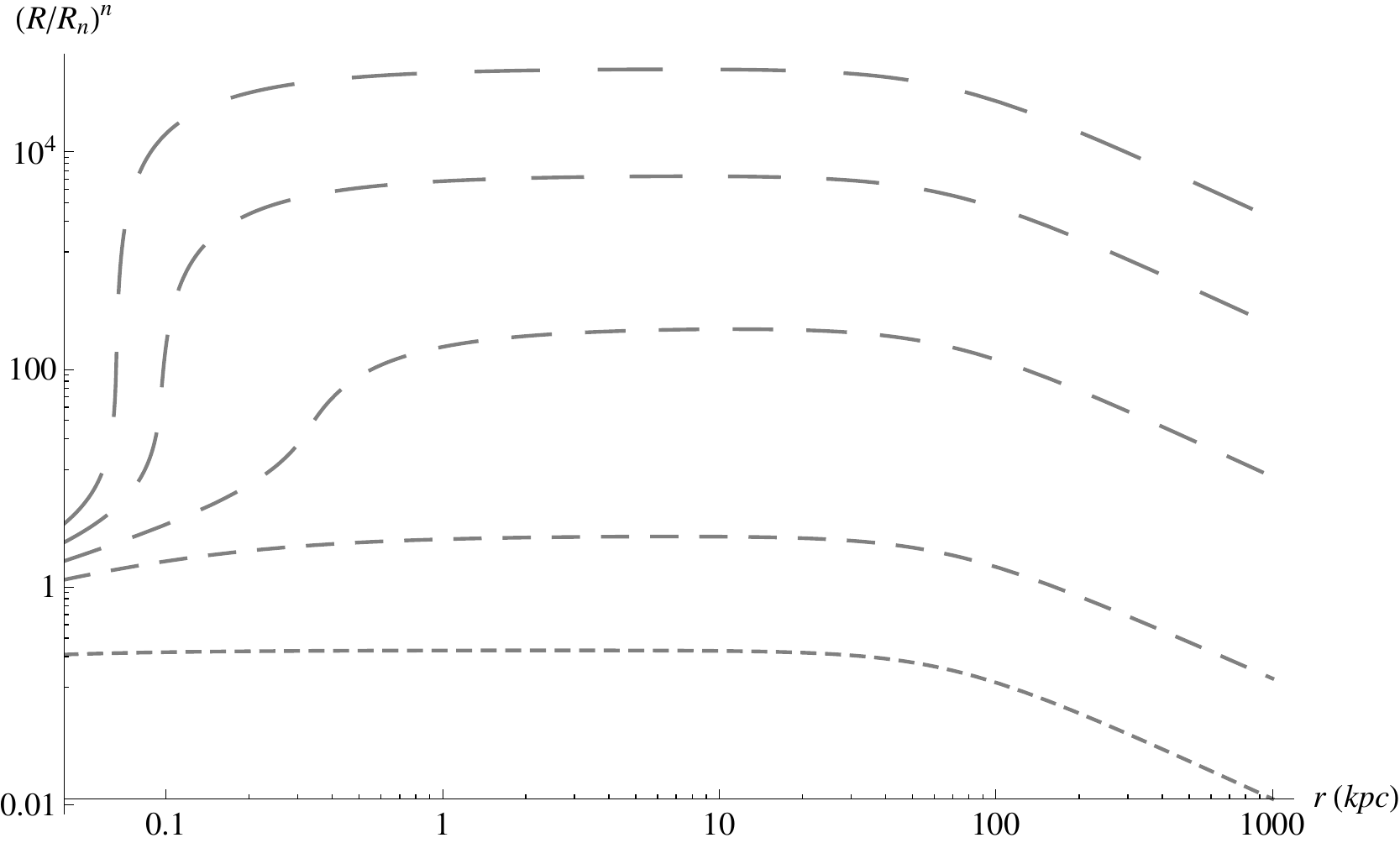}
  \caption{Different dark matter mimicked density profiles for different values of the characteristic length scale, $r_n=1.00$, $6.31$, $10.00$, $15.84$, and $25.12\;{\rm Mpc}$, where the dashing increases with the value of $r_n$, compared with the observed dark matter component (black). These values are obtained for the best fit value $n=0.43$.}
  \label{r0fig}
\end{figure}

Numerically, one finds that the mimicked dark matter profile is almost independent of the characteristic length scale $r_n$, provided one sets the bound $r_n \lesssim 1\;{\rm Mpc}$. This independence is easily explained: a different $r_n$ induces a change on the boundary conditions of Eq. (\ref{rhovar}), which translates into a change of the dimensionless solution $\varrho$ (not shown); however, this is counteracted by the presence of $r_n$ in the relation between the latter and the mimicked dark matter density, Eq. (\ref{var1}).

Fig. (\ref{r0fig}) shows that values above $r_n \sim 1\;{\rm Mpc}$ lead to a mimicked dark matter density that strongly deviates from the observed profile, with a sharp rise (and a subsequent fall into the expected outer slope). The onset of this rise decreases with higher values of $r_n$ --- {\it i.e.}, lower values of $R_n$.

Despite the complexity of the differential Eq. (\ref{rhovar}), the existence of an upper bound for this independence seems to be due to the fact that, since the exponent $n$ is positive in the considered range, a lower $R_n = 1/r_n^2$ translates into an increased value of $f_2(R)$: as $r_n$ approaches $r_0 \approx 12.8~{\rm Mpc}$ from below, one eventually gets $(R/R_n)^n \sim 1$ (in the inner, higher curvature region), in contradiction with the assumption of a perturbative non-minimal coupling.

\begin{figure}[tbp]
  \centering
  \includegraphics[width=\columnwidth]{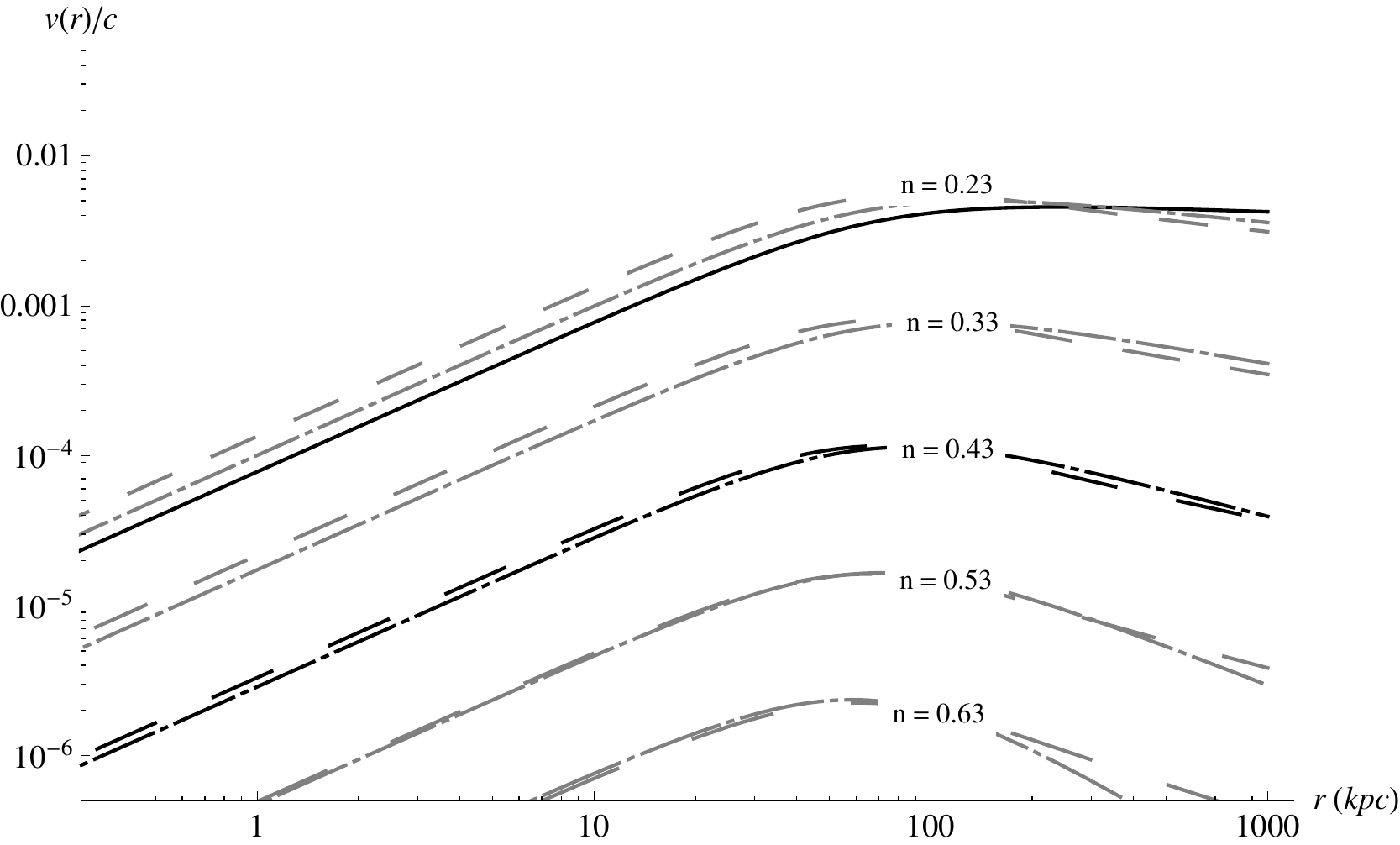}
  \caption{Additional radial velocity resulting from the non-geodesic motion due to the dark matter component (dashed) and to the mimicked dark matter (dot-dashed) for different values of $n$, and also for the total density profile (black).}
  \label{extra_forces}
\end{figure}

\begin{table*} 
 \centering
\begin{tabular}{c|cc|cc|ccc} 
Cluster &$\alpha$&$\beta$&  $n_{in}$ & $n_{out}$ & $r_{0.2} (10^{-10} {\rm kpc})$& $\Delta M/M$  & $\sigma_{\log}^2$\\ \hline
$\text{A133}$ & $0.996$ & $0.575$ & $0.502$ & $0.425$ & $2.70$ & $0.187$ & $0.143$ \\
$\text{A262}$ & $1.674$ & $0.333$ & $0.163$ & $0.667$ & $0.31$ & $1.744$ & $0.032$ \\
$\text{A383}$ & $2.018$ & $0.583$ & $-0.009$ & $1.001$ & $5.30$ & $1.325$ & $0.031$ \\
$\text{A478}$ & $1.493$ & $0.715$ & $0.253$ & $0.285$ & $0.17$ & $0.940$ & $0.373$ \\
$\text{A907}$ & $1.554$ & $0.594$ & $0.223$ & $0.406$ & $4.30$ & $0.076$ & $0.096$ \\
$\text{A1413}$ & $1.217$ & $0.651$ & $0.392$ & $0.349$ & $0.74$ & $2.195$ & $0.228$ \\
$\text{A1795}$ & $1.06$ & $0.545$ & $0.470$ & $0.455$ & $1.20$ & $1.166$ & $0.194$ \\
$\text{A1991}$ & $1.516$ & $0.501$ & $0.242$ & $0.499$ & $0.06$ & $1.022$ & $0.088$ \\
$\text{A2029}$ & $1.131$ & $0.539$ & $0.435$ & $0.461$ & $0.095$ & $5.238$ & $0.816$ \\
$\text{A2390}$ & $1.917$ & $0.696$ & $0.041$ & $0.304$ & $4.30$ & $0.348$ & $0.106$ \\
$\text{RX J1159+5531}$ & $1.762$ & $1.215$ & $0.119$ & $-0.215$ & $2.90$ & $0.178$ & $0.234$ \\
$\text{MKW 4}$ & $1.628$ & $1.224$ & $0.186$ & $-0.224$ & $0.002$ & $2.518$ & $0.024$ \\
$\text{USGC S152}$ & $2.644$ & $0.453$ & $-0.322$ & $0.547$ & $0.27$ & $1.087$ & $0.041$ \\
$\text{A586}$ &  &  & &  & 5.8 & $1.023$ & $0.207$
\end{tabular}
\caption{Cluster parameters, predicted inner and outer exponents (Eqs. (\ref{n1}) and (\ref{n2})), and quantities obtained for the best-fit scenario with fixed $n= 0.2$: relative mass difference $\Delta M/M$, $\sigma_{\log}^2$ measure (Eqs. (\ref{mass_deviation}) and (\ref{sigma2}), respectively) and maximum value of the characteristic length scale $r_{0.2}$ compatible with a perturbative extra force (Eq. (\ref{extra-force})).}
\label{clusters_values}
\end{table*}

\subsubsection{Non-geodesic motion}

\begin{figure}[tbp]
  \centering
  \includegraphics[width=\columnwidth]{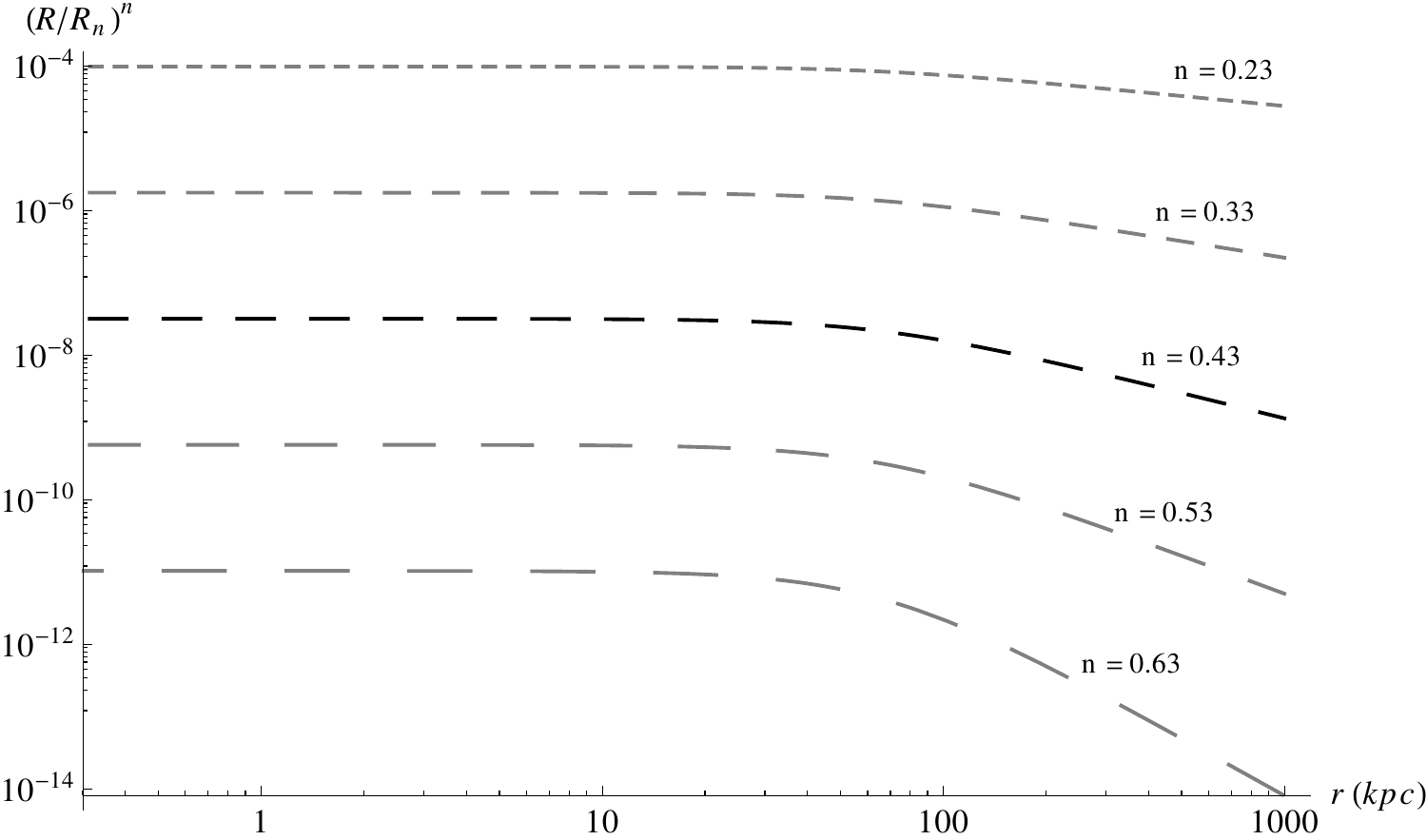}
  \caption{Resulting power-law couplings, $f_2(R)=\left(R/R_n\right)^n$, after the integration of the differential equation, 
for different values of $n$.}
  \label{couplings}
\end{figure}

In Fig. (\ref{extra_forces}) one presents the plot of the additional radial velocity $\De v = \sqrt{ r f^r}$ arising from non-geodesic motion as a function of distance to the cluster's center, for the considered range of values for $n$ and a fixed $r_n = 10^{-2}~{\rm pc}  $. Notice that, despite the fact that the observable quantity is the velocity dispersion $\sigma$, estimated from the velocities of individual clusters, these curves enable one to see the relevance of the extra force that appears due to the presence of the non-minimal coupling.

Since the scalar curvature ({\it i.e.} the total density) is decreasing, one concludes that the related force counteracts the gravitational attraction, as discussed in Sec. \ref{subsectionnon}. Thus, stability requires that this additional velocity be much smaller than the Newtonian radial velocity profile: for a fixed value of $r_n$, this is attained for higher values of $n$. In particular, one concludes that a power-law coupling with $n=0.43$ can effectively mimic the dark matter component of the Abell cluster A586, with a negligible effect on the velocity profile $v(r)$.

As already discussed in Sec. \ref{nongeodesicalmotion}, one expects a decrease of the effect of the extra force for increasing exponent $n$ (for fixed $r_n$). Indeed, as Fig. (\ref{couplings}) depicts, the considered range for $n$ always leads to a perturbative $f_2(R)$, for a fixed $r_n= 10^{-2}~{\rm pc}$. However, values of the exponent in the range $n \lesssim 0.23$, which imply $b( 10^{-2}~{\rm pc}  , n ) > 1$, do not enable a sufficiently strong suppression of the extra force: although the additional (outward) radial velocity is indeed smaller than $c$, it may still be above the Newtonian counterpart --- the converse would require one to compute the condition \eq{lower_bound} for $n= 0.23$, in order to obtain a bound on $r_{0.23}$.

\begin{figure*}
  \centering
  \includegraphics[width=\columnwidth]{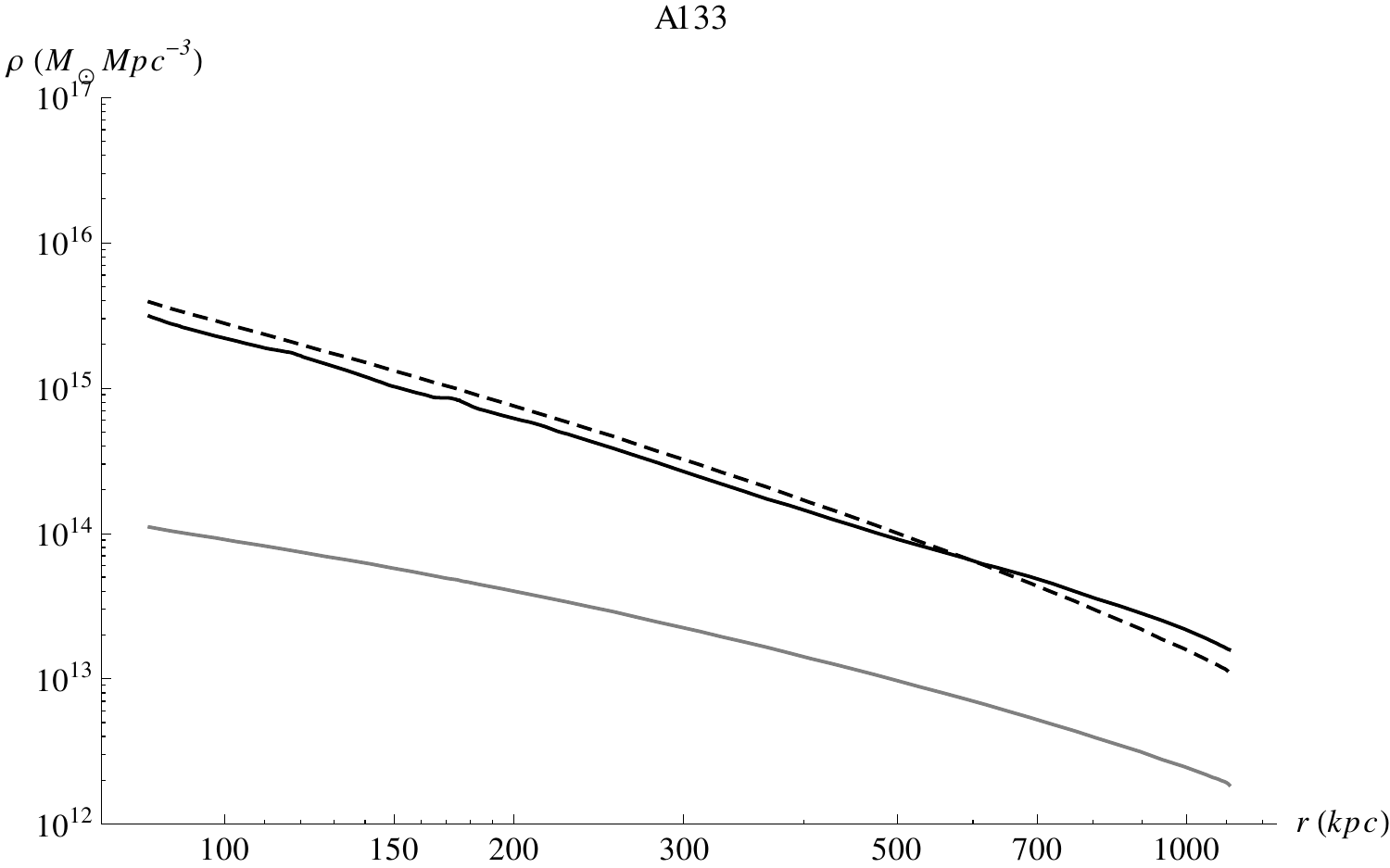}\hspace*{0.2cm}
  \includegraphics[width=\columnwidth]{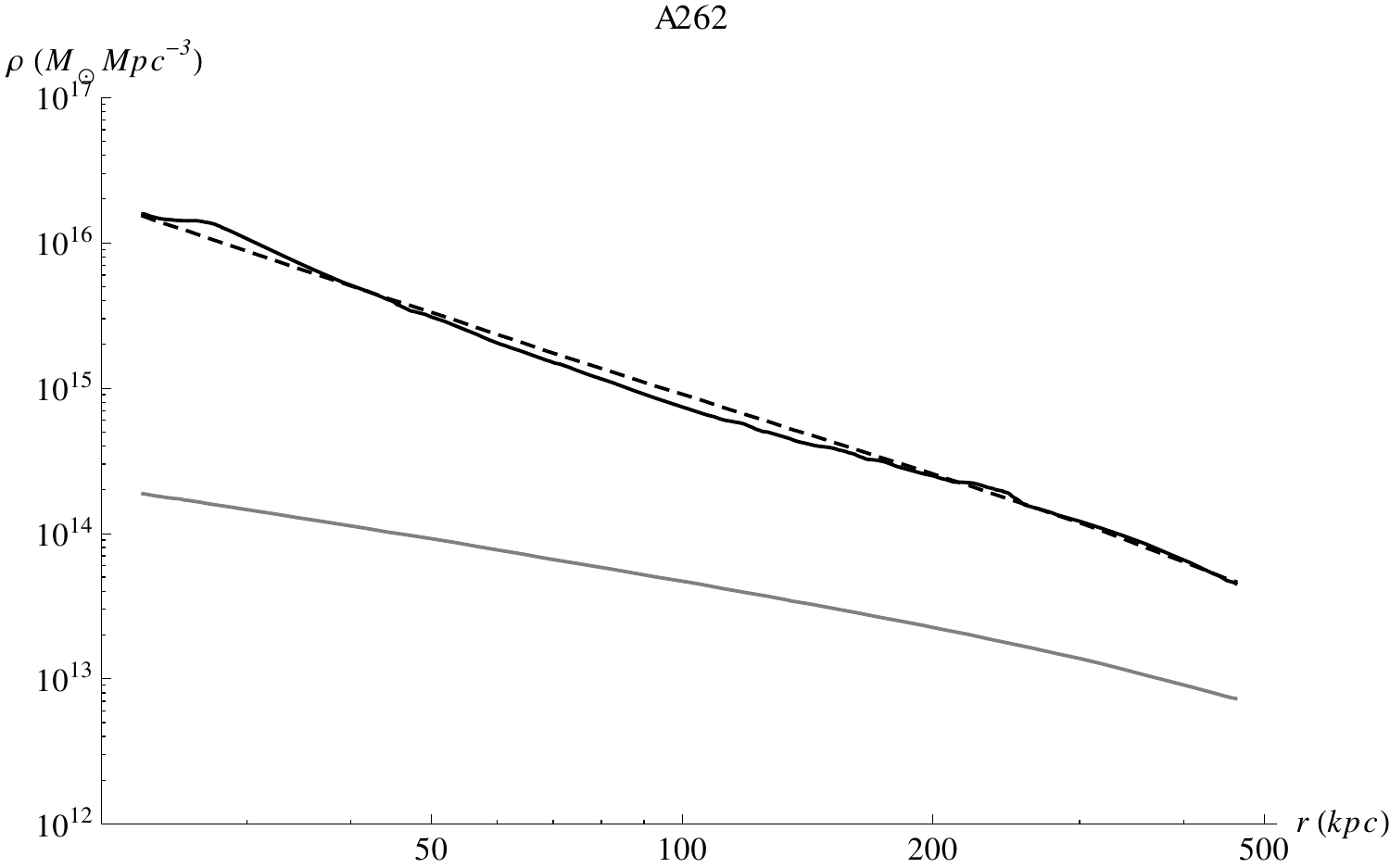}\\
  \includegraphics[width=\columnwidth]{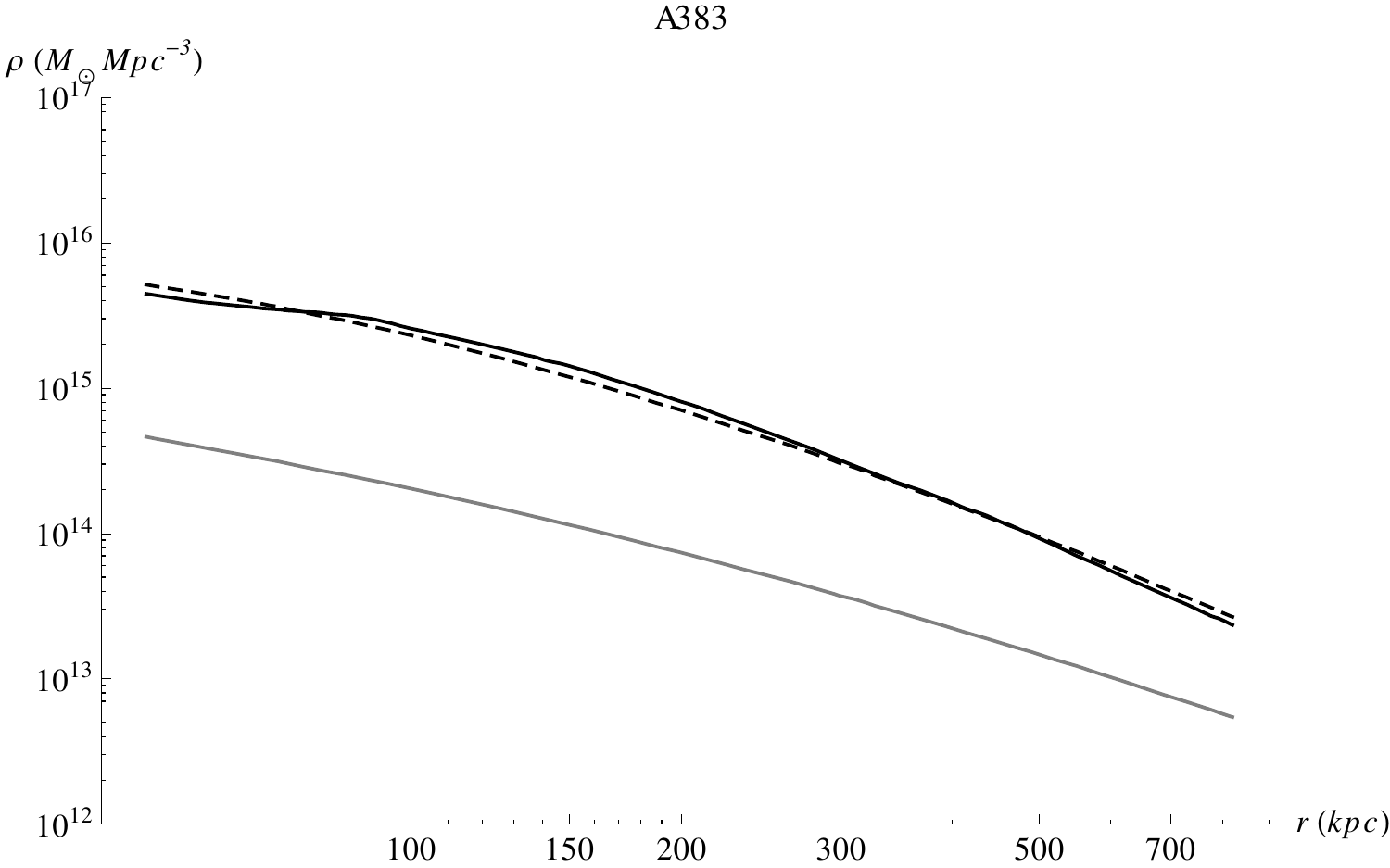}\hspace*{0.2cm}
  \includegraphics[width=\columnwidth]{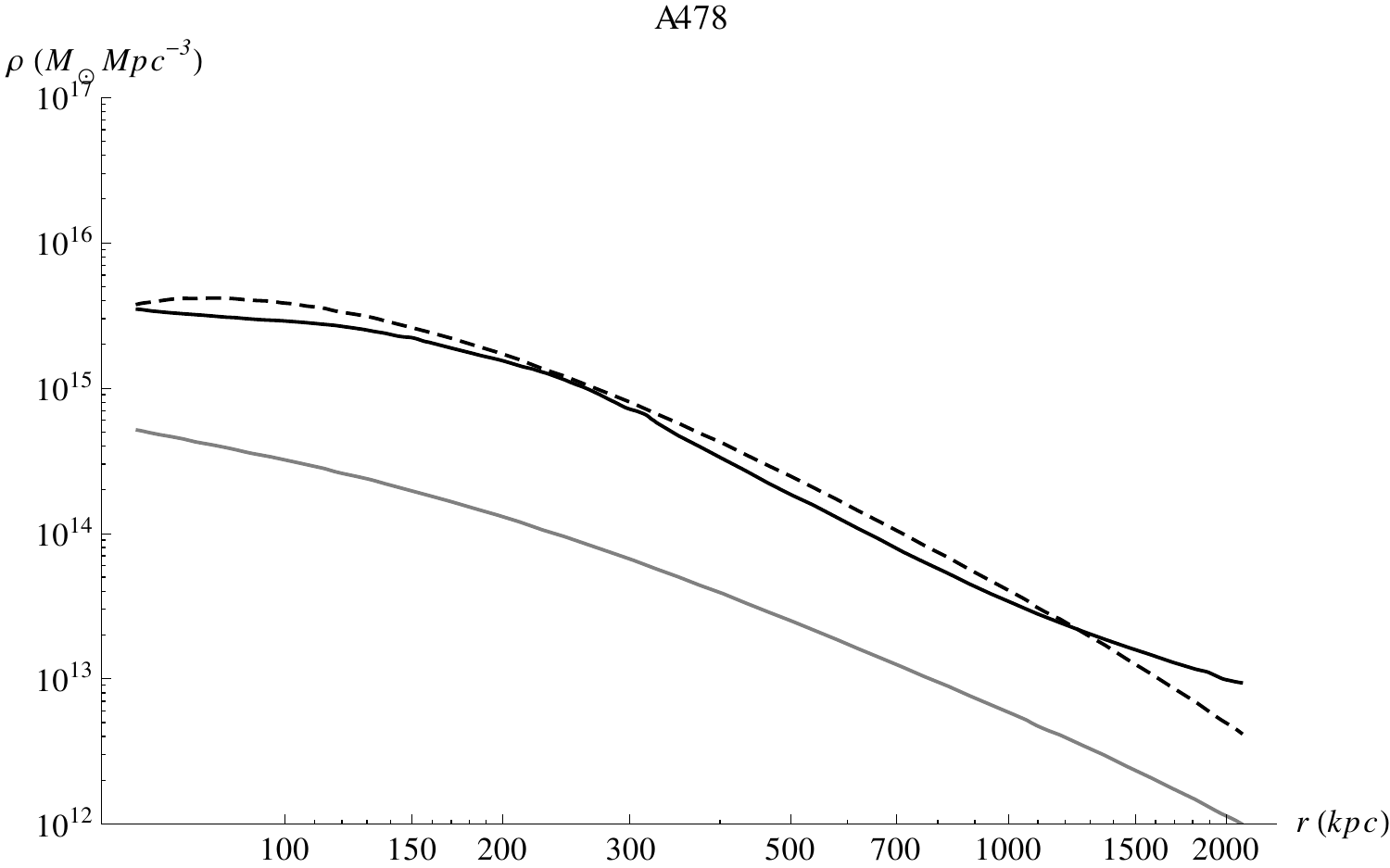}\\
  \includegraphics[width=\columnwidth]{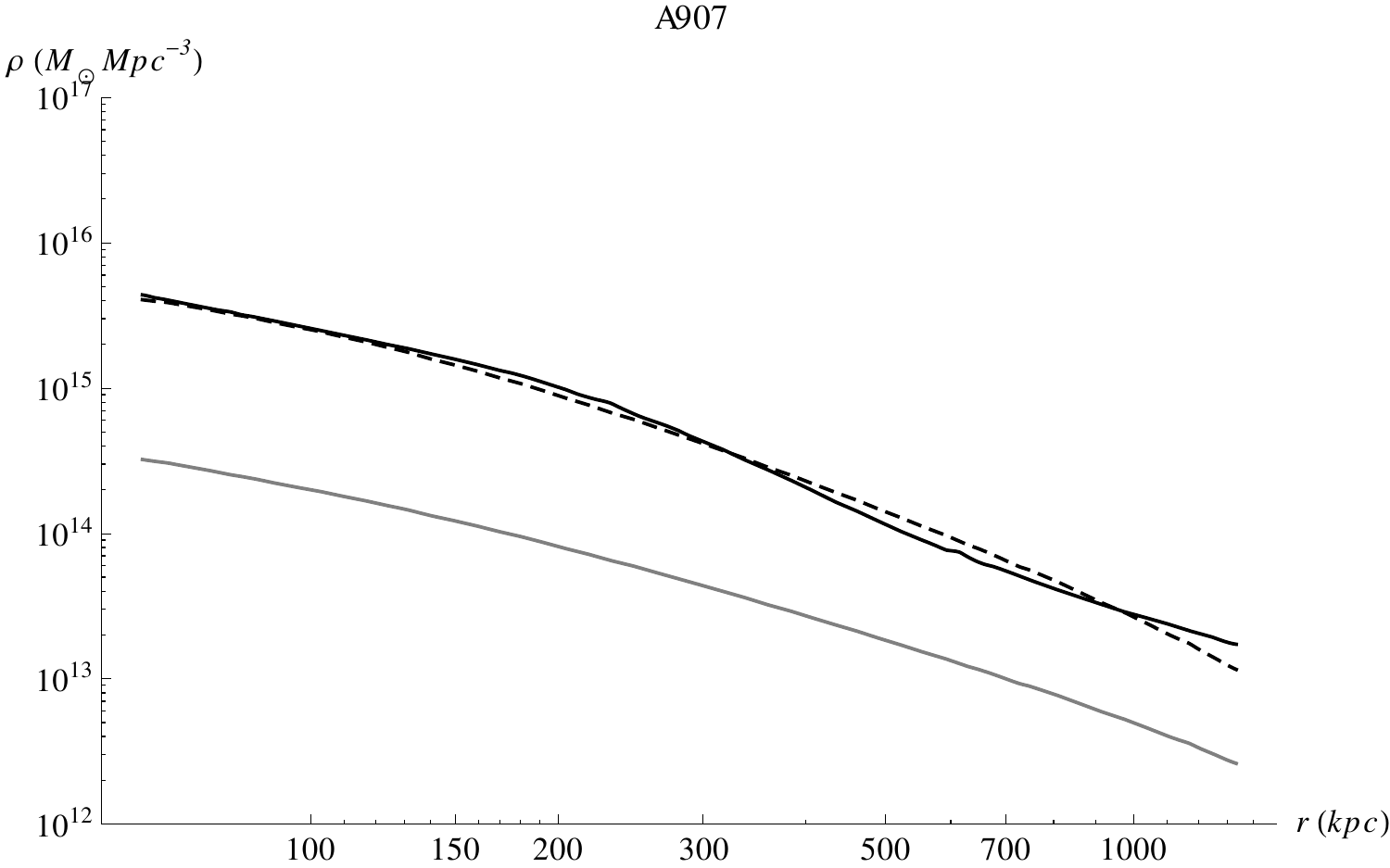}\hspace*{0.2cm}
  \includegraphics[width=\columnwidth]{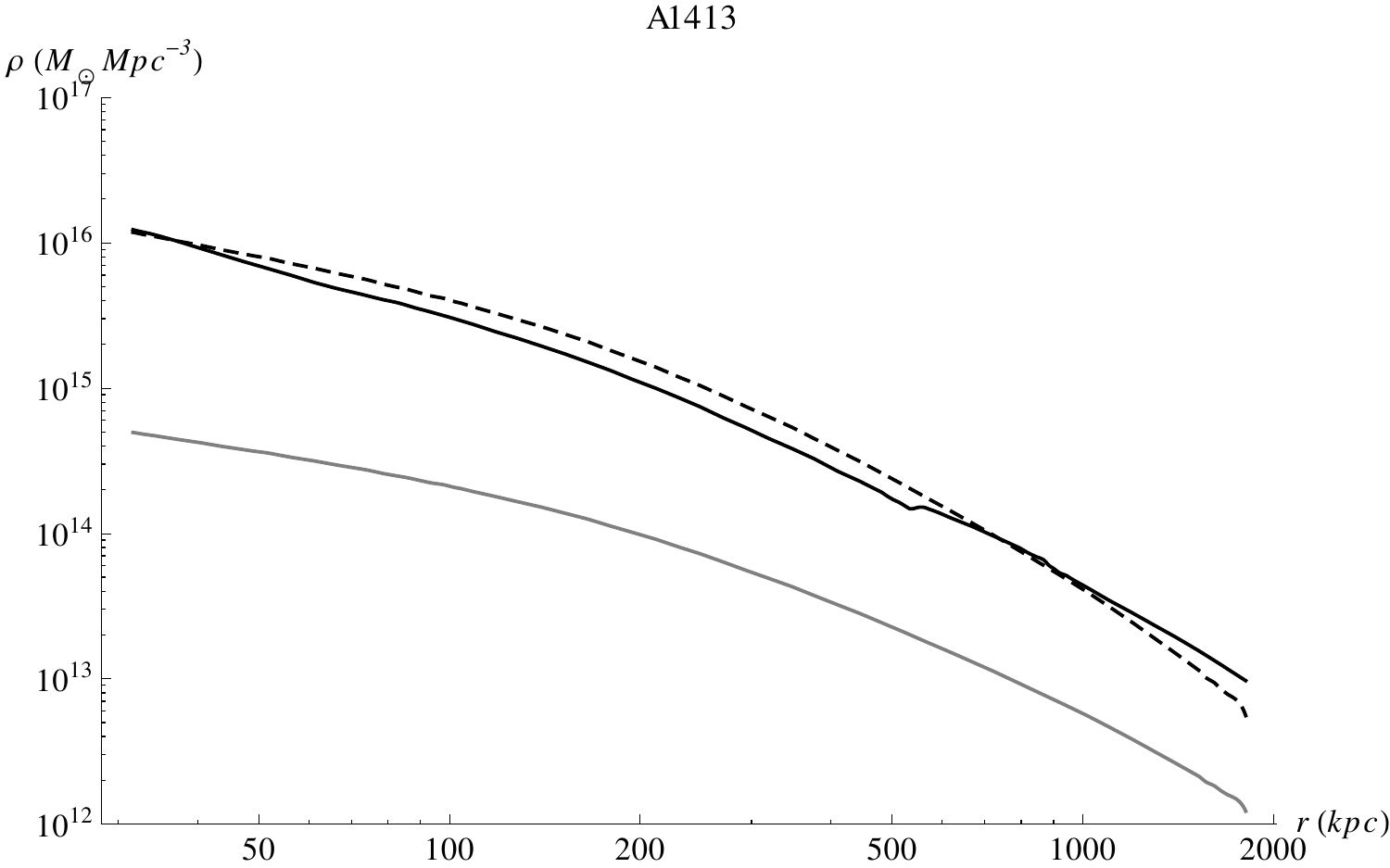}
\caption{The obtained dark matter mimicked profile using a power-law function fixed at $n=0.2$ (dashed), along with the density profiles for dark matter (black) and visible (gray) components for each cluster.}
  \label{mimic1}
\end{figure*}

Using \eq{defdm0} and the relevant values obtained from Ref. \cite{A586_data}, one has $\ep = 0.3626$ and $\rho_{dm0} \approx 10^{16} M_\odot / {\rm Mpc}^3 \rightarrow r_0 \approx 12.8~{\rm Mpc}$. Inserting this into \eq{lower_bound}, together with $r_c = 67~{\rm kpc}$ and $r_n =  10^{-2}~{\rm pc} $, one finds that the extra force is smaller than the Newtonian force when $b( 10^{-2}~{\rm pc}  , n ) < 1$, which is equivalent to the constraint, $n > 0.246$, and should be compared with the aforementioned lower bound $n > 0.23$, obtained from Fig. \ref{extra_forces}.

The lack of sensitivity of the mimicked dark matter density profile on the value of $r_n $ may be used to establish a more stringent upper bound on this parameter. To do so, one fixes the best fit value $n = 0.43$ and searches for the value of $r_n$ that fulfills the condition $b(r_n, 0.43) < 1$, which is equivalent to the constraint $r_n < 51~{\rm pc}$.
   
\section{Chandra Cluster Sample}

One now extends the dark matter mimicking mechanism to a large sample of galaxy clusters, aiming to obtain an encompassing description of their ``dark matter'' content. For this purpose, one uses a sample of nearby relaxed clusters derived from the high quality data of {\it Chandra} \cite{vikhlinin}. 

As already seen, in the case of Abell cluster A586, one resorts to the scaling $\rho_{dm}\sim\rho_g^{1/(1-n)}$ to estimate the exponent $n$ of the power-law non-minimal coupling. However, since the density profiles are now more evolved, alternatively one considers the behaviour of the inner and outer regions, that is one resorts to two different exponents, $n_{in}$ and $n_{out}$, respectively. Indeed, the dark matter density profile follows a NFW profile given by
\beq
\rho_{t}\sim \left(r \over r_s \right)^{-1}\left(1+{r \over r_s} \right)^{-2}~~,
\eeq

\noindent such that for small radii, $\rho_t\sim (r/r_s)^{-1}$, and for outer regions, $\rho_t\sim (r/r_s)^{-3}$; the gas density profile follows a generalized NFW model 
\cite{vikhlinin},

\beq
\rho^2_g\sim n_0^2{(r/r_c)^{-\alpha}\over (1+r^2/r_c^2)^{3\beta-\alpha/2}}~~, \label{generalized_beta}
\eeq

\noindent for which, $\rho_g\sim (r/r_c)^{-\alpha}$, in inner regions, and $\rho_g\sim (r/r_c)^{-3\beta}$ for outer regions. Thus, for small radii one has, from Eq. (\ref{n-estimative}), 

\beq
{1 \over 1-n_{in}}={1\over \alpha/2}\quad\rightarrow\quad n_{in}=1-{\alpha\over 2}~~,
\label{n1}
\eeq

\noindent and, for large radii,

\beq
{1 \over 1-n_{out}}={3\over 3\beta}\quad\rightarrow\quad n_{out}=1-\beta~~.
\label{n2}
\eeq

\noindent The results of  Eqs. (\ref{n1}) and (\ref{n2}) for each cluster are shown in Table \ref{clusters_values} using the reported values for the $\alpha$ and $\beta$ parameters \cite{vikhlinin}. Since $n$ is a parameter of the coupling function Eq. (\ref{powerlawcoupling}), it should be universal: thus, one performs a simultaneous fit of the ``dark matter'' component of all clusters, with a fixed $n$. The characteristic length scale $r_n$ is varied individually for each cluster, allowing for the analysis of the perturbative regime of the non-minimal coupling Eq. (\ref{powerlawcoupling}) and the extra force Eq. (\ref{extra-force}), discussed below --- following the approach detailed for the Abell A586 cluster in the previous section.

\begin{figure*}
  \centering
  \includegraphics[width=\columnwidth]{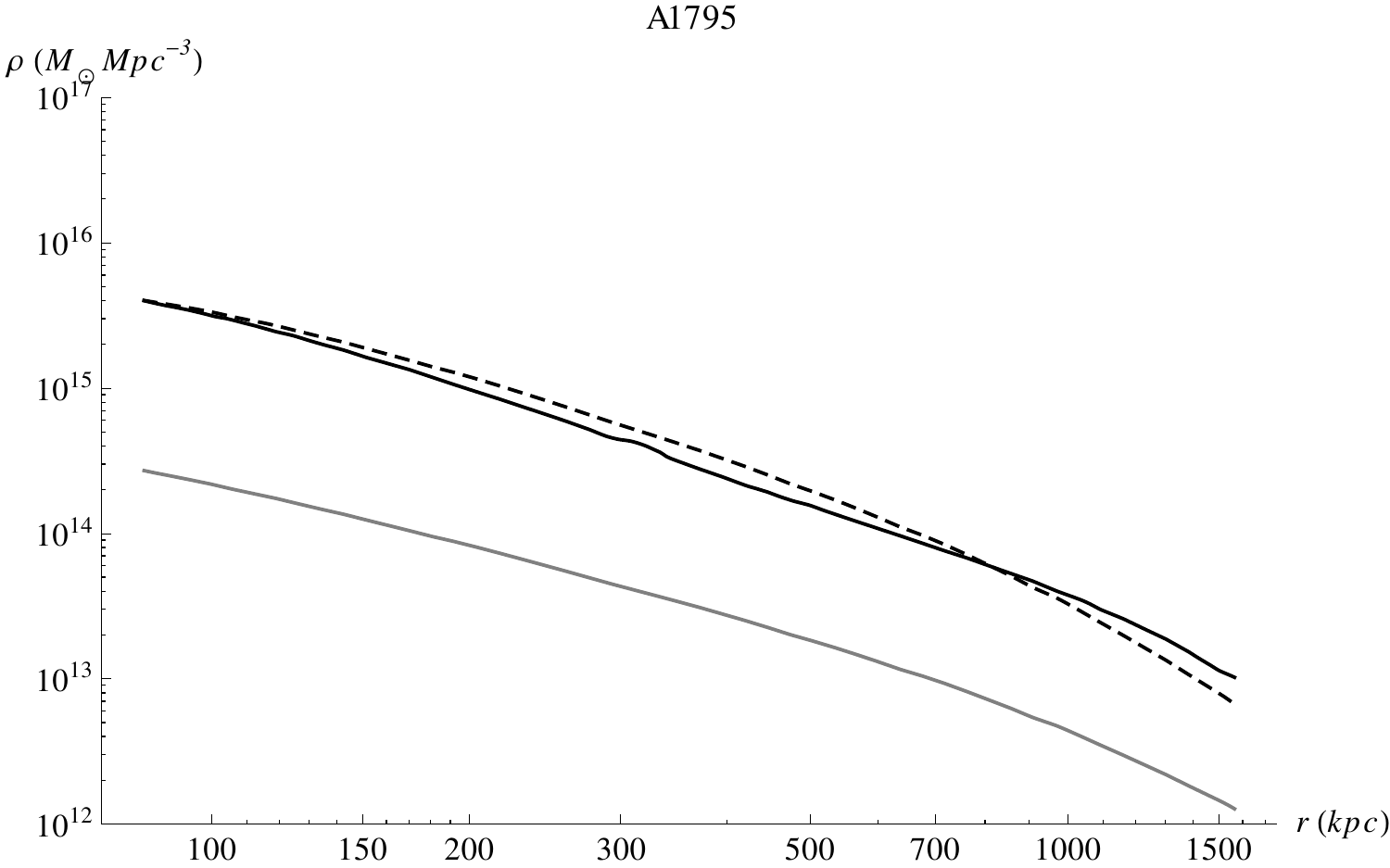} \hspace*{0.2cm}
  \includegraphics[width=\columnwidth]{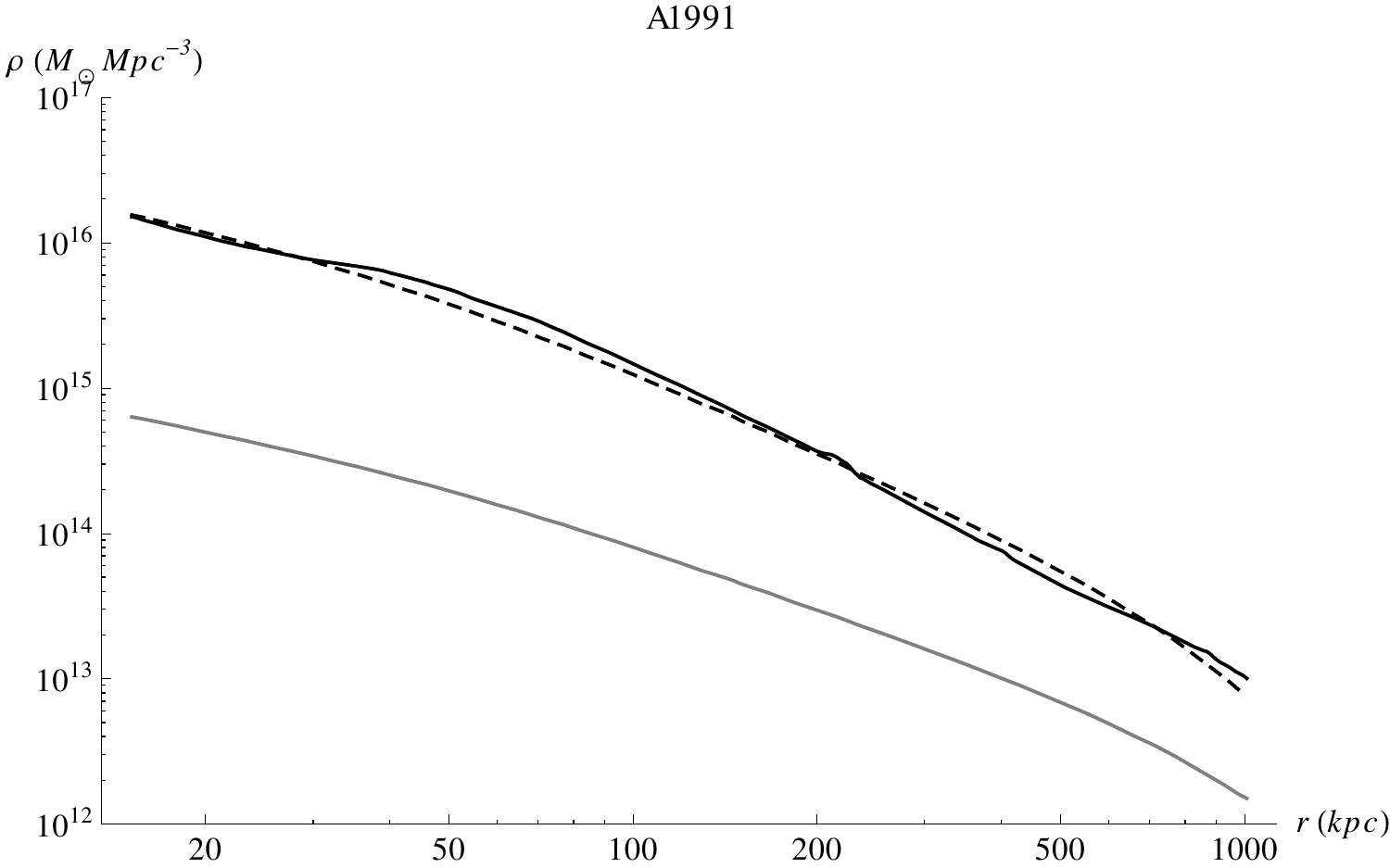}\\
  \includegraphics[width=\columnwidth]{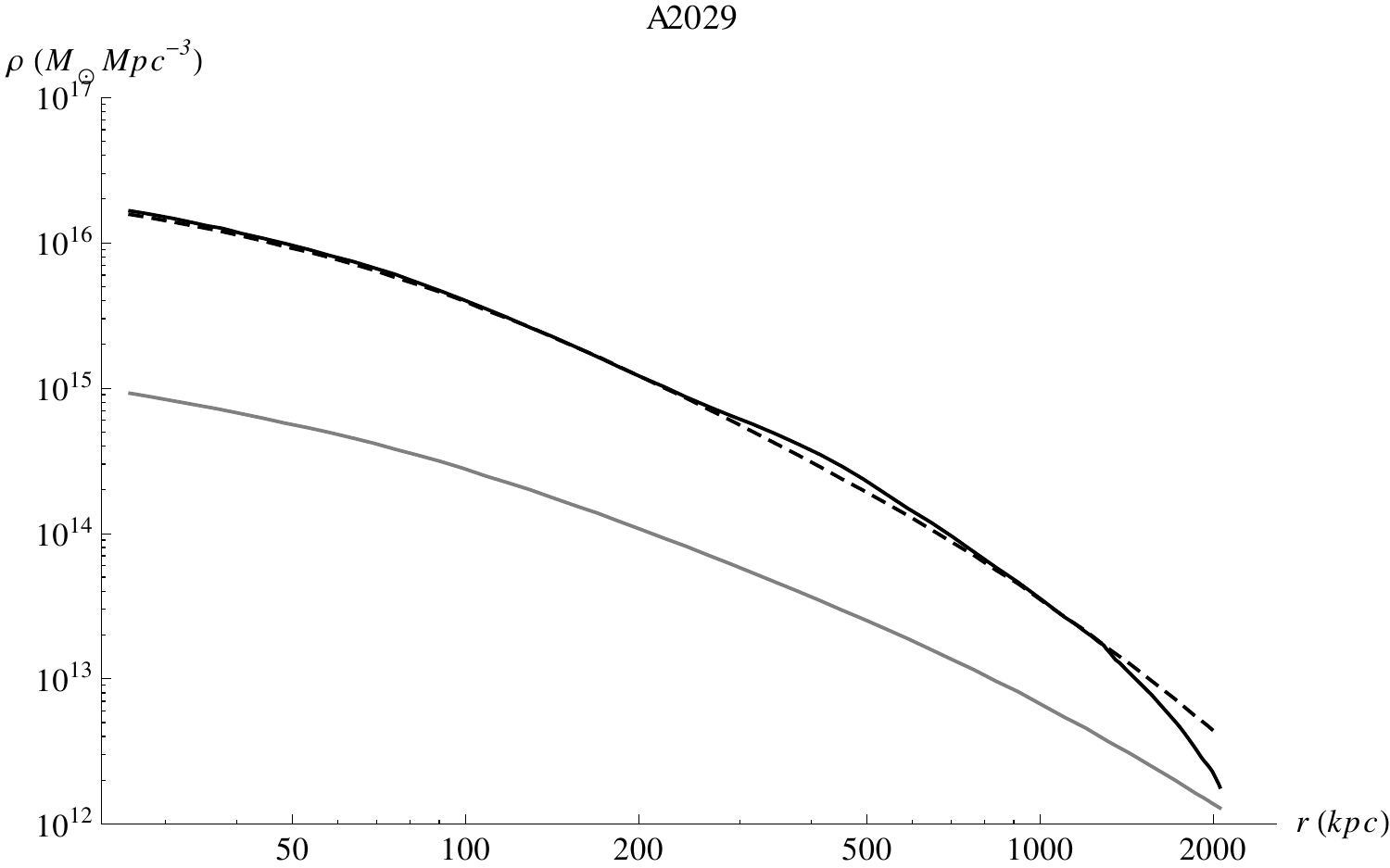}\hspace*{0.2cm}
  \includegraphics[width=\columnwidth]{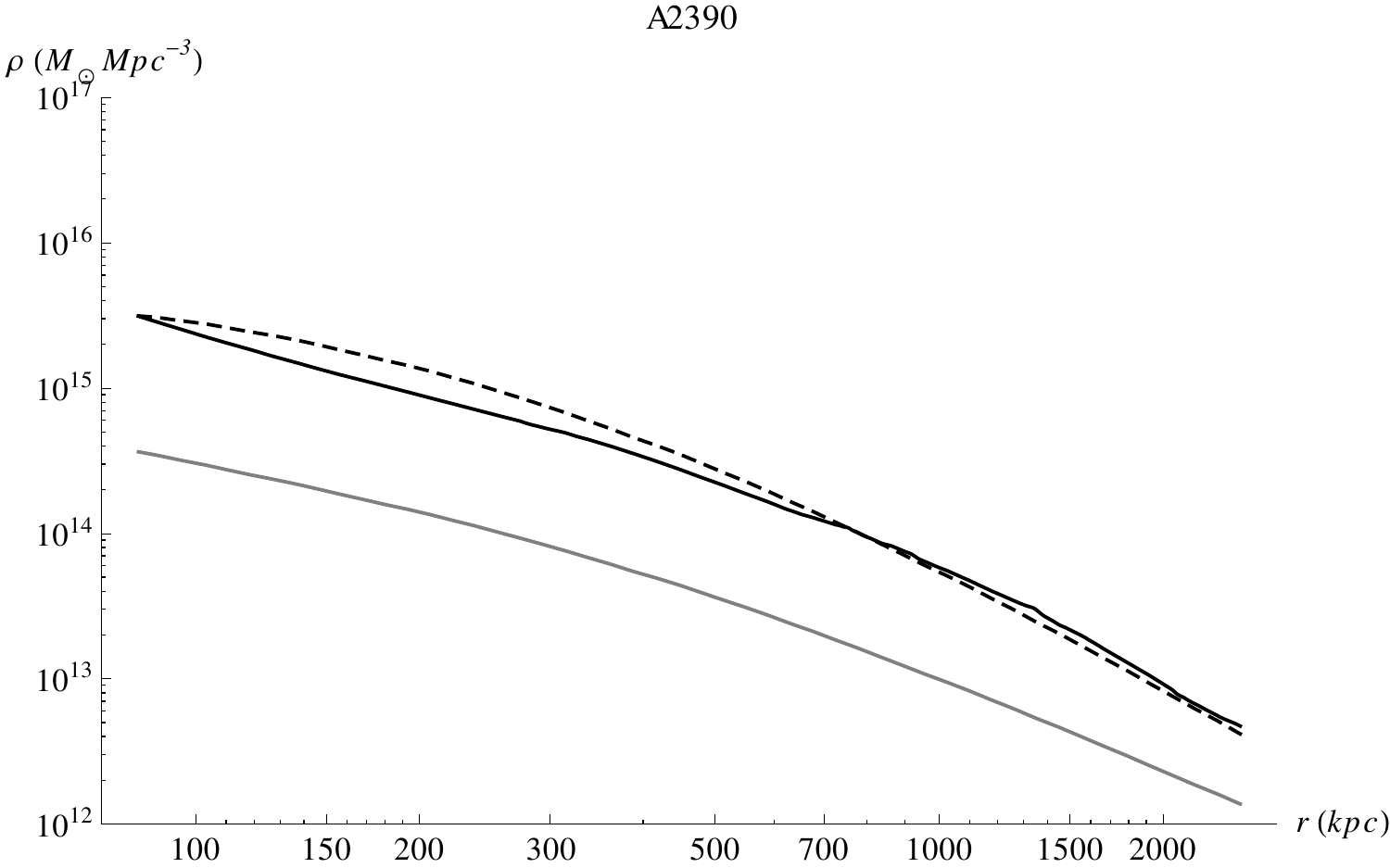}\\
  \includegraphics[width=\columnwidth]{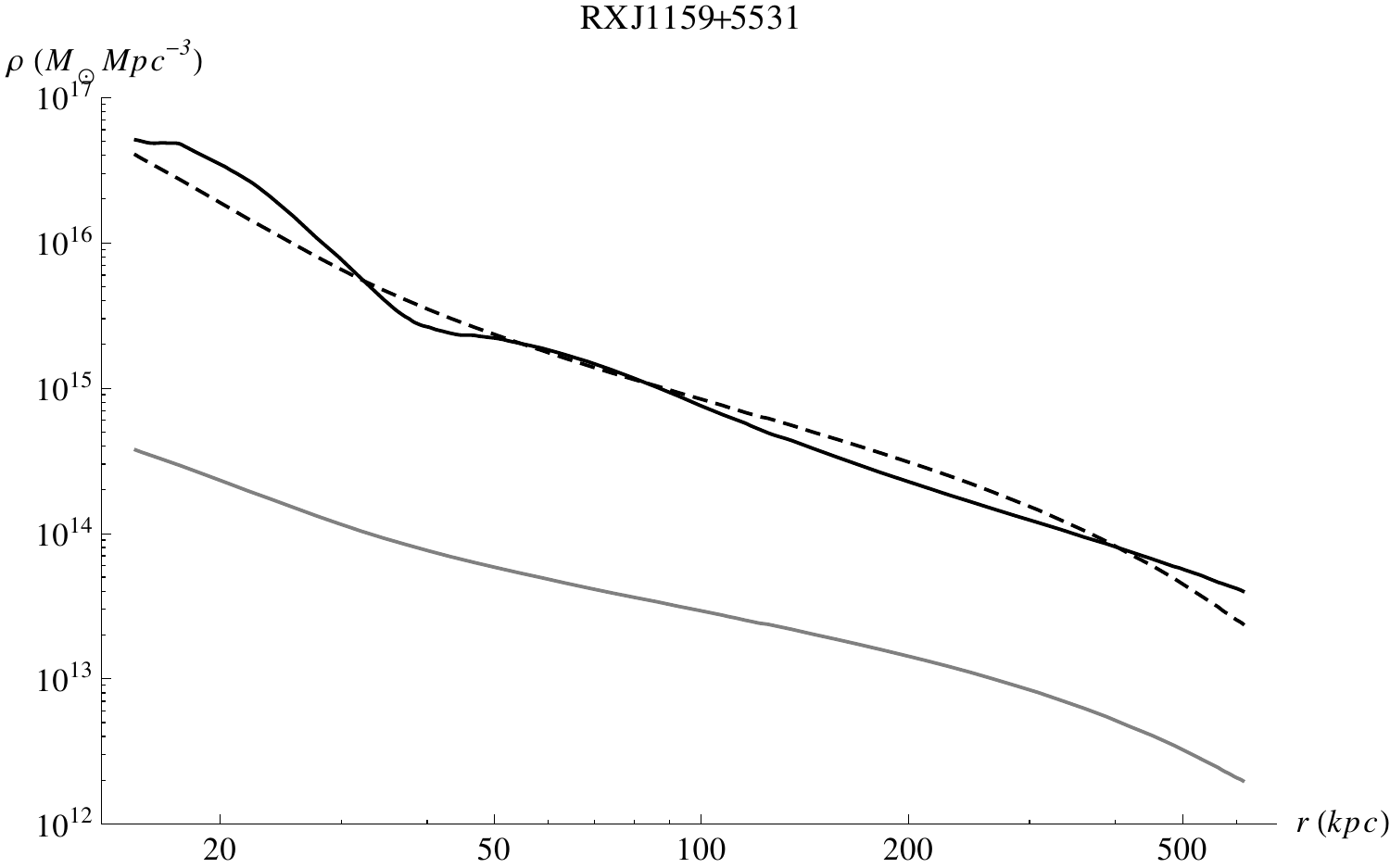}\hspace*{0.2cm}
  \includegraphics[width=\columnwidth]{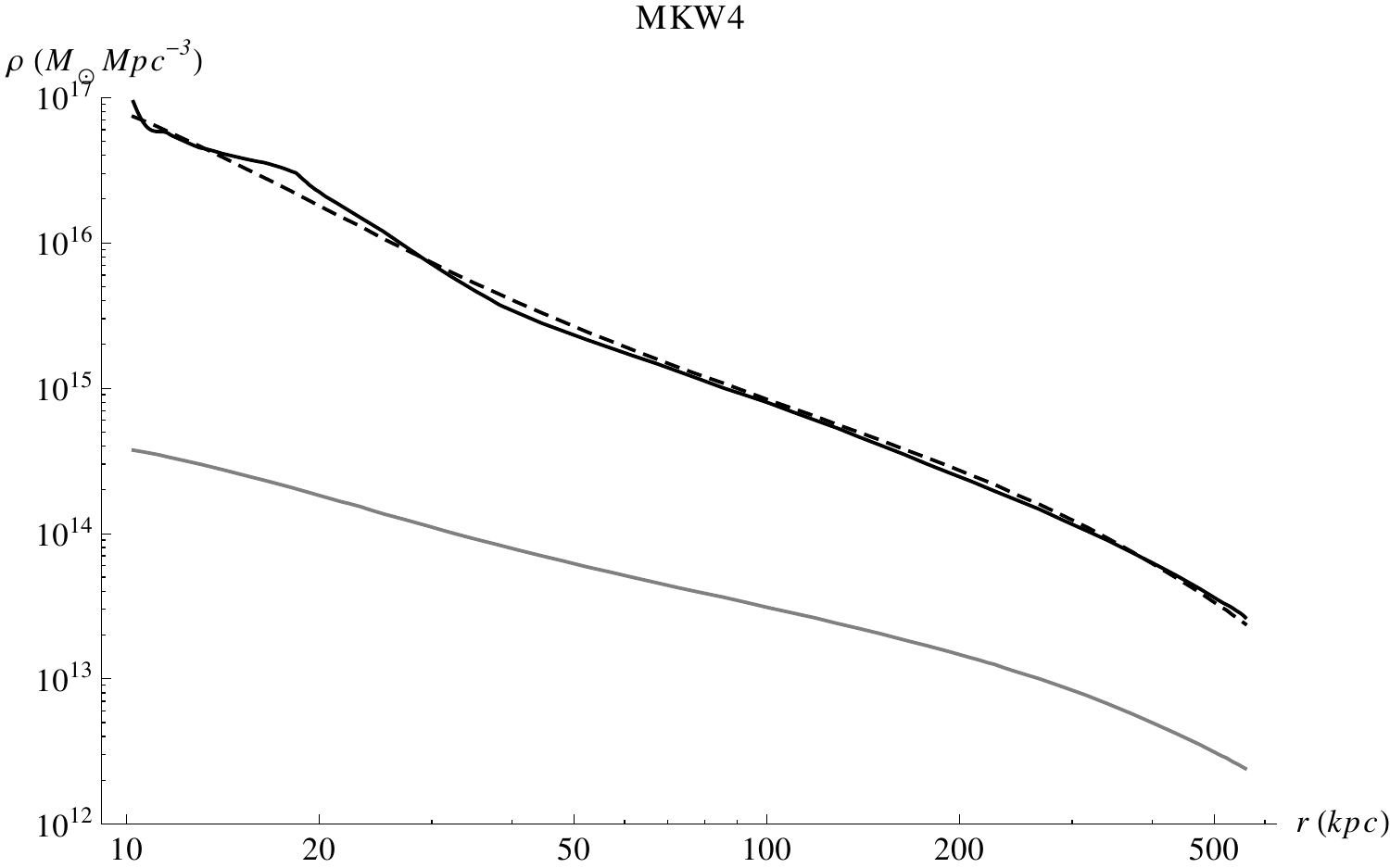}\\
  \includegraphics[width=\columnwidth]{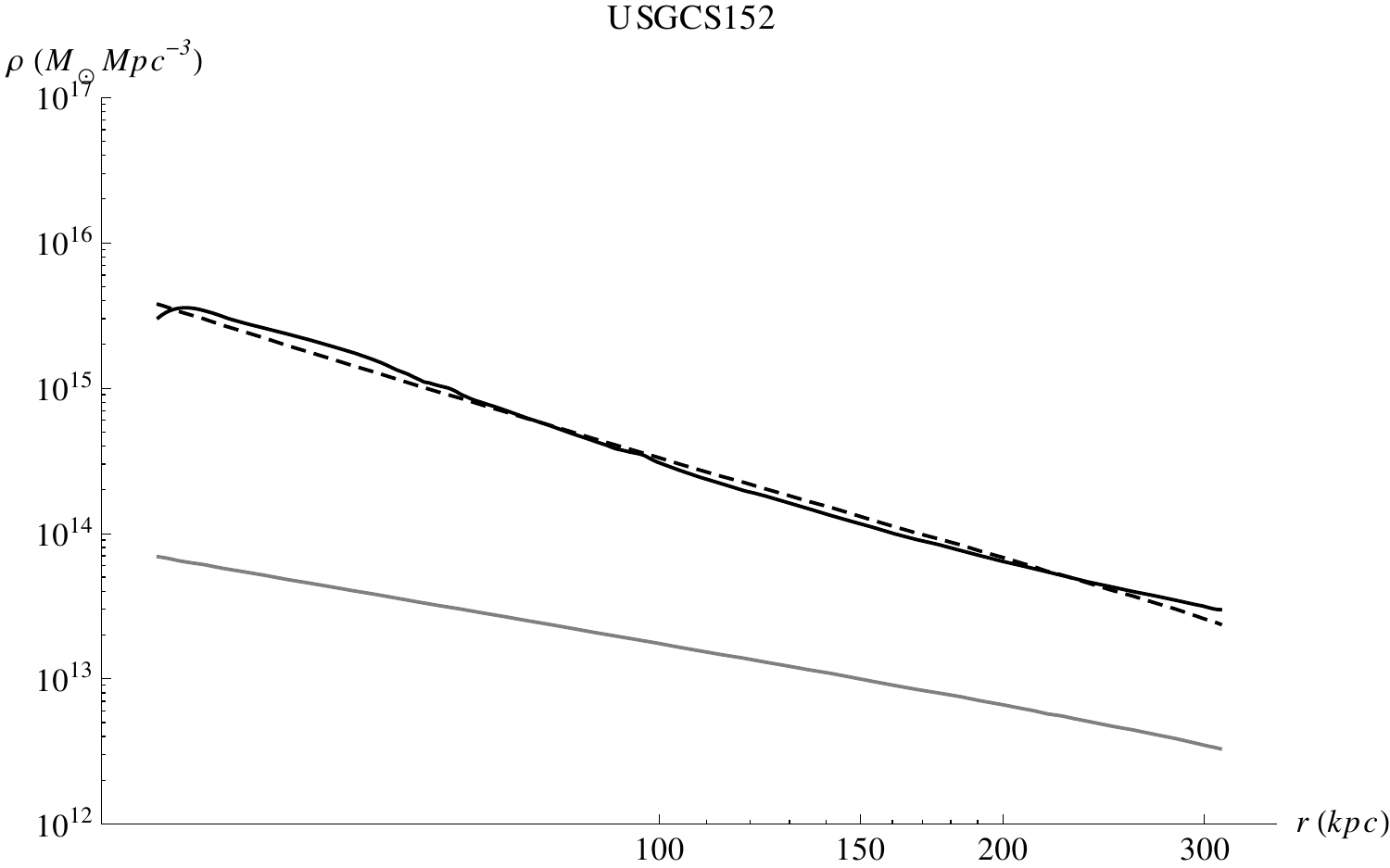}\hspace*{0.2cm}
  \includegraphics[width=\columnwidth]{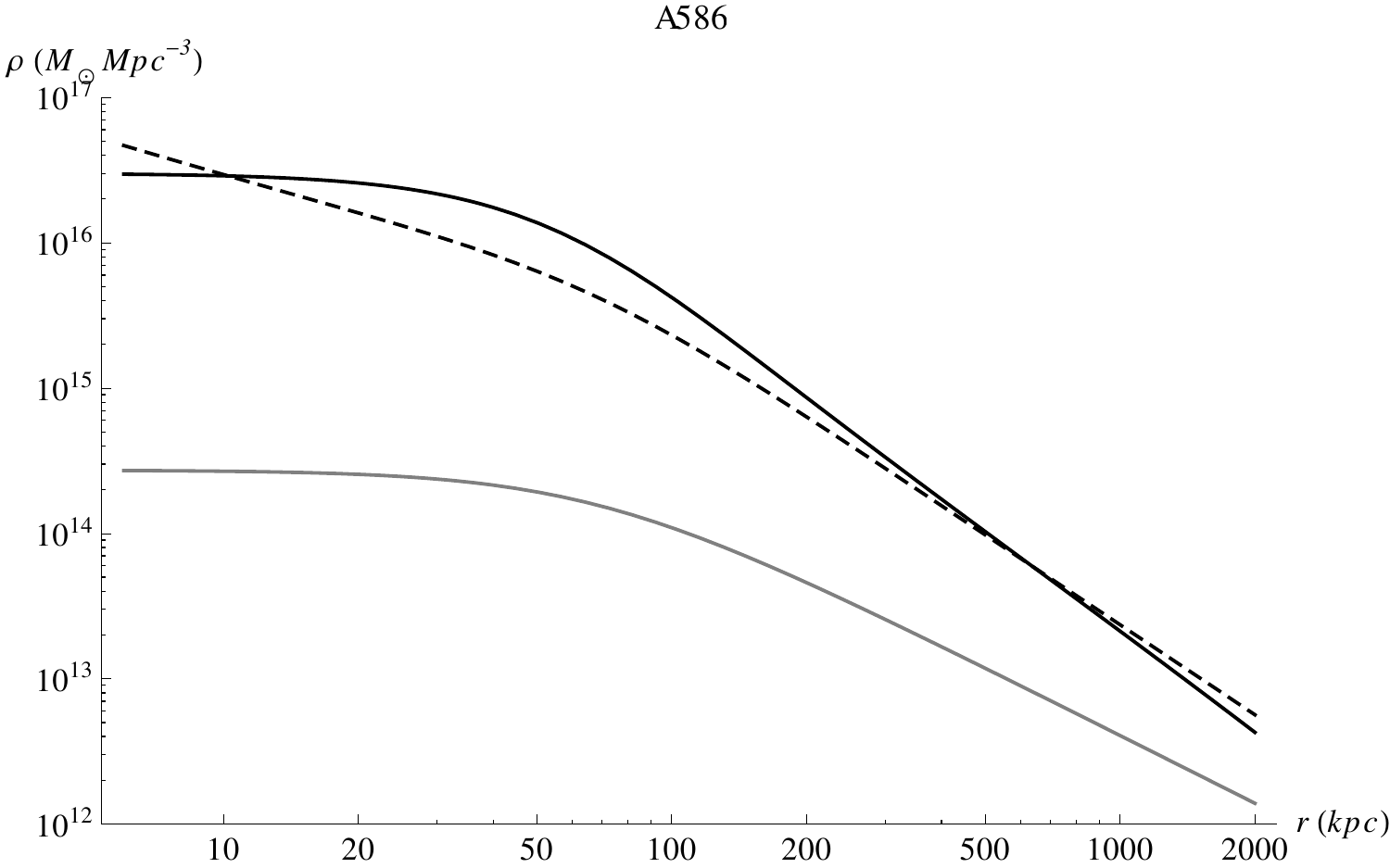}
  \caption{Same as Fig. (\ref{mimic1}) except for different clusters. The last graph corresponds to the Abell A586 cluster. }
  \label{mimic2}
\end{figure*}

Numerically, one finds that the best fit occurs for $n= 0.2$; the initial conditions ({\it i.e.} the initial value for $\varrho$ and its derivative) are varied within one order of magnitude, reflecting the uncertainty on the dark matter distributions reported in Ref. \cite{vikhlinin}. The resulting dark matter profiles yield the  dashed curves presented in Figs. (\ref{mimic1}) and (\ref{mimic2}): these minimize the difference between the observed and mimicked dark matter components, leading to small values of the estimator $\sigma^2_{log}$, Eq. (\ref{sigma2}), and mass difference $\Delta M/M$, Eq. (\ref{mass_deviation}), as can be seen in Table \ref{clusters_values}. 

As in the case of Abell cluster A586, one finds that the mimicked dark matter profile is almost independent of the characteristic length scale $r_{0.2}$, as long as the non-minimal coupling Eq. (\ref{powerlawcoupling}) is perturbative, $f_2(R) \ll 1$. This leads to the bound $r_{0.2}\lesssim 0.1~{\rm kpc}$, which is approximately independent of the cluster. As before, a more stringent bound, $r_n \lesssim 10^{-13}~{\rm kpc}$, is obtained from the requirement that the extra force Eq. (\ref{extra-force}) is smaller than the Newtonian force.

The independence of the obtained mimicked dark matter distributions on the value of $r_n$ again emerges because a different value for the latter leads to a change in the boundary conditions of Eq. (\ref{rhovar}), which implies a change of the dimensionless solution $\varrho$; this is compensated by the presence of $r_n$ in the relation between the latter and the mimicked dark matter density, Eq. (\ref{var1}).

\subsection{A586 cluster global fit results}

Notice that the fitting of the Abell A586 cluster dark matter profile using the universal exponent $n=0.2$ (final graph of Fig. \ref{mimic2}) leads to the values $\Delta M/M=1.023\%$, and $\sigma_{\log}^2=0.207$ --- naturally higher than those obtained from the $n=0.43$ best fit of this cluster alone.

There is a fundamental qualitative difference between the density profiles derived for the A586 cluster and the set of 13 Chandra clusters: the later do not exhibit a transition between an inner region with approximately constant density and an outer region with decreasing density of approximately constant slope (as embodied in the $\be$-model \eq{visible_profile}).

This drives the ensuing distinction between the fits for the A586 cluster obtained in this study: indeed, as Fig. \ref{densities} shows, the best fit $n=0.43$ obtained by considering only this cluster yielded an initial plateau for the mimicked dark matter density. This is to be expected, as most of the mass comes from the integration of the density in this higher, approximately constant region --- thus leading to a reduced mass difference $\De M/M$.

On the contrary, the mimicked dark matter profile of this cluster obtained in the global fit (final graph of Fig. \ref{mimic2}) does not have an approximately flat density for the inner region, but compensates by more closely following the observed slope of the outer regime --- hence the proximity between the obtained best fit parameter $n=0.2$ and the value obtained from the scaling law $\rho_{dm(n)} \sim \rho^{1/(1-n)}$, $n=0.279$.

This occurs because the minimization procedure of the global fit is mostly driven by the remaining thirteen Chandra clusters: as it turns out, the best fit for this set (excluding the A586 cluster) is $n \sim 0.1$ (not shown in this study); a balanced fitting solution with all fourteen clusters is thus obtained by relaxing the constraint of a flattened inner density region for the mimicked dark matter profile of the A586 cluster.

\section{Relevance in other contexts}

The non-minimal coupling considered here may be regarded as an approximation to a more evolved form for $f_2(R)$, which would encompass couplings considered in previous studies, which perhaps hints at a Laurent series,
\beq f_2(R) =  \sum_{n = - \infty}^{\infty} \left({R \over R_n}\right)^n ~~. \label{Laurent} \eeq

\noindent Each term would be valid in a particular regime: early {\it vs.} late time, central {\it vs.} long range, galactic {\it vs.} cluster size, {\it etc}. In particular, this series should also include the aforementioned $n=-1$ and $n=-1/3$ terms, relevant to mimic dark matter at the galactic level, as well as a linear coupling $R/R_1$ considered in the context of preheating and inflationary dynamics; the corresponding parameters are 

\beqa 
r_{-1} &\equiv& \left(R_{-1}\right)^{-1/2} = 21.5~{\rm Gpc}~~, \\ \nonumber 
r_{-1/3} &\equiv& \left(R_{-1/3}\right)^{-1/2} = 1.69 \times 10^6~{\rm Gpc}~~, \\ \nonumber 
r_1 &\equiv& R_1^{-1/2} = 4.8 \sqrt{\xi} \times 10^{-29}~{\rm m} ~~, 
\eeqa

\noindent with $10 < \xi < 10^4$ \cite{coupling_inflation}.

Since the relevant coupling terms at galactic scales have negative exponents, it is trivial to conclude that their effect is negligible during inflation, when the curvature is much higher. Conversely, the effect of the terms involved in preheating and inflation are irrelevant at astrophysical or cosmological scales at present.

Similarly, the cosmological effect of the best fit coupling for the considered sample of clusters (with $n=0.2$) may be disregarded, as it is of order

\beq \left({R_{cosmo}\over R_n}\right)^n \sim \left({r_n\over r_H}\right)^{2n} \lesssim \left({10^{-13} ~{\rm kpc}\over 4.2~{\rm Gpc}}\right)^{0.2} \sim 10^{-4} ~~.\eeq

\noindent where $r_H = c /H = 4.2~{\rm Gpc}$ is the Hubble radius.

Thus, one is left with the possible conflict between astrophysical relevant terms. If one considers three power-law couplings with $n=-1$, $-1/3$, and $0.2$, the resulting dark matter profile will not be a sum of the individual contributions, due to the non-linearity of \eq{second_equation}. Even if one disregards the resulting interaction between terms (a reasonable approximation, as shown in Ref. \cite{mimic}), the exponents simply yield the slope for each derived dark matter contribution, not for their relative values. Indeed, when numerically integrating \eq{rhovar}, one imposes boundary conditions on the overall dark matter density ({\it i.e.} the dimensionless function $\varrho$) --- the relative contributions of each dark matter component are then determined dynamically.

Given the scaling relation \eq{implicit}, a larger $n < 1$ implies a steeper dark matter component $\rho_{dm} \sim \rho^{1/(1-n)}$. Resorting to the $\rho \sim r^{-3\be}$ behaviour of the visible mass density Eq. (\ref{generalized_beta}) and using the average value $\be \approx 0.653  $, this translates into

\beqa n=-1:~\rho_{dm} \sim r^{-0.98}~~&,&~~n=-{1 \over 3}:~\rho_{dm} \sim r^{-1.5}~~~~,\nonumber \\  n=0.2&:&~\rho_{dm} \sim r^{-2.2}~~~~.\eeqa

\noindent Thus, the $n= 0.2$ merely gives rise to a quickly decaying component: the $n=-1$ and $n=-1/3$ contributions eventually dominate at large distances, even if the former dominates at short ones.

In order for the contributions from the $n = -1$ and $-1/3$ to be neglected, it suffices that they are subdominant at cluster scales, but dominant at the long range of galactic scales. One does not pursue an explicit computation of the effect of these three power-law terms here, given that the main scope was to describe clusters. Nevertheless, it is not difficult to see that the difference between the galactic and cluster typical densities do allow for a consistent disentangling of the effects of these two sets of terms.

\section{Conclusions and Outlook}
In this work, the possibility of mimicking the dark matter component of a galaxy cluster in the context of a model, \eq{model}, where the scalar curvature is non-minimally coupled with matter is discussed. For that, one first assesses the viability of the model using the relaxed Abell A586 cluster, given the availability of quality data for most of its relevant dynamical parameters.

Using a power-law non-minimal coupling $f_2(R) = (R/R_n)^n$, one obtains the best fit for $n=0.43$, with a strong constraint on the typical length scale $r_n \equiv 1/\sqrt{R_n} < 51~{\rm pc}$ arising from the requirement that the induced extra-force is smaller than the Newtonian one. A weaker constraint $r_n \lesssim r_0 = 12.8 ~{\rm Mpc}$ arises from the requirement that the coupling is perturbative, $f_2(R) \ll 1$.

The obtained fit does not correspond to the analytical value for $n$ computed from the scaling relation between dark and visible matter, as the latter would give rise to a mimicked dark matter density that does not follow the observational curve (although its outer behaviour exhibits the same slope). Instead, a shift from this analytical value is enforced, whereas the mimicked density is slightly steeper than the observed profile, but with a negligible difference in total mass.

Furthermore, the exponent $n$ is large enough to keep the non-minimal coupling perturbative, with $f_2(R) \sim 10^{-8}$. This suppresses the destabilizing effect of the outward extra-force arising from the non-conservation of the energy-momentum tensor, as previously found in the context of galaxy rotation curves (where two negative exponents $n=-1$ and $n=-1/3$ were shown to correspond to the best fit for the rotation curves) \cite{mimic}.

Subsequently, this dark matter mimicking mechanism was successfully extended to a large sample of galaxy clusters using the available profiles for the visible matter density and temperature obtained from modifications of the usual $\beta$-model, Eq. (\ref{visible_profile}), and the polytropic temperature profile, Eq. (\ref{temperature_profile})  \cite{vikhlinin}. For this purpose, one has used the same power-law coupling function, obtaining an exponent $n=0.2$ from a simultaneous fit to all considered clusters. As in the detailed case of the Abell A586 cluster, the fitting procedure is also independent of the length scale $r_n$, as long as the coupling is perturbative, $f_2(R) \ll 1$. By the same token, a more stringent upper bound of $r_n\lesssim 10^{-13}~{\rm kpc}$ is obtained from the requirement that the extra force due to non-geodesic motion is much smaller than the Newtonian force.  

The fitting procedure depends on the underlying model \eq{model}, but also on several other assumptions --- namely, sphericity, virialization, and thermal equilibrium. Clearly, deviations from any of these hypotheses will introduce an undetermined bias into the best fit scenario: the choice of the Abell 586 cluster as a initial test case was motivated by the wish to minimize these sources of error, as this cluster is very well characterized and has a good spherical morphology \cite{A586_data}. 

In galaxy clusters with more evolved structures, one may observe deviations due to the presence of non-sphericity and non-thermal pressure support, and the density profile may be much rougher than in well-behaved clusters. However, an adequate description of the mimicked dark matter content was obtained for clusters with more evolved internal structures, namely RXJ1159+5531, MKW 4 and USGC S152, thus supporting the universality and robustness of the proposed mimicking mechanism.

Studies addressing different scenarios have resorted to power-law couplings with different exponents than the best fit value $n= 0.2$ here reported --- which may be regarded as another term in a putative series expansion of a more general non-minimal coupling. The specific form of the coupling function can only be obtained from its consideration in different scenarios, in order to reconstitute its full functional dependence on the Ricci scalar curvature. Thus, the relevance of each term in the several scenarios considered in previous works was also discussed. By the same reasoning, yet unprobed terms of this series may be charted through the study of other phenomena and environments (ranging from astrophysical to cosmological scales), where distinct curvatures and densities are at play.

\end{document}